\documentclass[pra,aps,twocolumn,a4paper,showpacs]{revtex4}

\usepackage{amsmath} \usepackage{amsfonts} \usepackage{amssymb}
\usepackage{graphicx} \usepackage{epsfig}

\begin{document}

\title{A Single Atom Mirror for 1D Atomic Lattice Gases}

\author{A. Micheli and P. Zoller}

\affiliation{Institute for Theoretical Physics, University of
  Innsbruck, and\\
Institute for Quantum Optics and Quantum Information of the Austrian
Academy of Sciences, A-6020 Innsbruck, Austria}

\begin{abstract}
  We propose a scheme utilizing quantum interference to control the
  transport of atoms in a 1D optical lattice by a single impurity
  atom. The two internal state of the impurity represent a spin-1/2 (qubit), which in
  one spin state is perfectly transparent to the lattice gas, and in
  the other spin state acts as a single atom mirror, confining the
  lattice gas. This allows to ``amplify'' the state of the qubit, and
  provides a single-shot quantum non-demolition measurement of the
  state of the qubit.  We derive exact analytical expression for the
  scattering of a single atom by the impurity, and give approximate
  expressions for the dynamics a gas of many interacting bosonic of
  fermionic atoms.
\end{abstract}


\maketitle

\section{Introduction}\label{Sec:Introduction}

One of the fundamental models in quantum optics is the interaction of
a spin-$1/2$ system with a bosonic mode \cite{Cohen}. The most
prominent example is cavity quantum electrodynamics (CQED), where a
two level atom interacts with a single mode of the radiation field in
a high-Q cavity. CQED has been the topic of a series of seminal
experiments both in the microwave and optical regime, demonstrating
quantum control on the level of single atoms and photons in an open
quantum system \cite{Kimble,Haroche,Rempe,Walther}.

\begin{figure}[t]
  \begin{center}
    \includegraphics[width=.99\columnwidth]{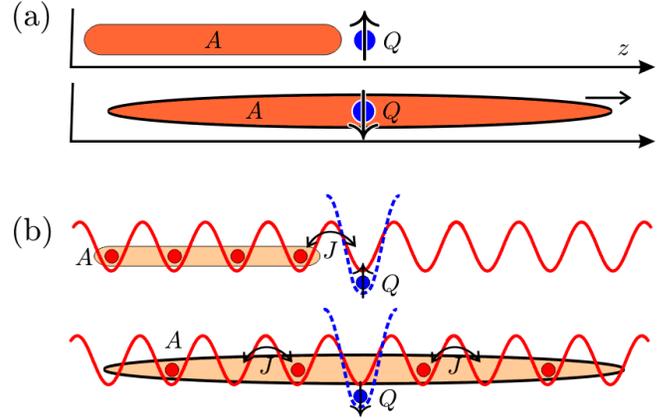}
    \caption{(a) A spin 1/2 impurity $Q$ used as a switch: in one spin
      state it is transparent to the probe atoms $A$, but in the other
      it acts as a single atom mirror. (b) Implementation of the SAT:
      The impurity atom $Q$ and the probe atoms $A$ are trapped
      separately in state-dependent 1D optical lattices. The probe
      atoms $A$ initially are in a Mott insulating
      state.}\label{Fig:setup}
  \end{center}
\end{figure}

In the present paper we will consider a system with the same basic
ingredients, however in the context of cold atoms and quantum
degenerate gases. The key feature of these systems is there
controllability and weak decoherence. In particular we employ two
aspects of control, the confinement of atoms in optical lattices
\cite{JakschZollerReview,Greiner,Mandel,Stoeferle} and (magnetic
or optical) Feshbach resonances as a way to manipulate atomic
interactions \cite{Bolda,Julienne,Holland,Theis}. According to the
setup described in Fig.~\ref{Fig:setup}(a) we will study the
dynamics of an atomic quantum gas in 1D (with a single internal
atomic state), representing bosonic or fermionic ``modes'',
controlled by an atomic spin-1/2 impurity. The quantum gas is
confined by tight trapping potentials (e.g. an optical or magnetic
trap), so that only the motional degrees along the $z$-axis in
Fig.~\ref{Fig:setup}(a) are relevant. In the $z$-direction the
motion is confined to the left by a trapping potential (e.g. a
blue sheet of light), while the atomic impurity restricts the
motion of the gas to the right due to collisional interactions of
the quantum gas with the impurity. The atom representing the
impurity can, for example, be a different atomic species in a
tight trapping potential, a configuration discussed in
Refs.~\cite{Recati,Raizen} as an {\em atomic quantum dot} ($0$D
system). Thus the impurity atom plays the role of ``single atom
mirror'' confining the quantum gas in an ``atomic cavity''.

\begin{figure}[t]
  \begin{center}
    \includegraphics[width=.99\columnwidth]{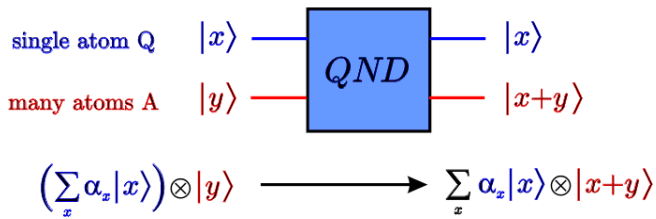}
    \caption{The single atom mirror as a macroscopic quantum
      gate. The qubit $Q$ entangles two distinguishable macroscopic
      phase of the probe atoms $A$ and provides for a quantum
      nondemolition interaction.}\label{Fig:qnd}
  \end{center}
\end{figure}

In our model system the impurity atom is an internal two level
system, which we write as an effective spin-1/2. In the following
we will also interpret this two-level system as a qubit with two
logical states $|0\rangle=|\downarrow\rangle$ and
$|1\rangle=|\uparrow\rangle$. Cold atom collision physics allows
for a situation where the collisional properties (scattering
length) of the impurity atom and atoms in the quantum gas are
spin-dependent. As illustrated in Fig.~\ref{Fig:setup}(a), we
assume that in one spin state, say $|\downarrow\rangle$, the single
impurity atom is completely transparent for the quantum gas, i.e.
the gas will leak out through the ``mirror''. In contrast, in the
other spin state the mirror atom is ``highly reflective''
confining the gas. For an impurity atom (qubit) initially prepared
in a spin superposition \[|\psi_Q(t=0)\rangle =
\alpha_\uparrow|\uparrow\rangle+\alpha_\downarrow|\downarrow\rangle\]
the combined system at a time $t$ will be in a macroscopic
superposition state
\begin{eqnarray}
  |\Psi(t)\rangle =
  \alpha_\uparrow|\uparrow\rangle|\phi_\uparrow(t)\rangle + \alpha_\downarrow|\downarrow\rangle|\phi_\downarrow(t)\rangle.\label{Eq:MacroCATstate}
\end{eqnarray}
with $|\phi_\sigma(t)\rangle$ many body wave functions of the gas
atoms. Thus $|\Psi(t)\rangle$ represents a Schr{\"o}dinger cat state
of two {\em entangled quantum phases} of gas atoms, the first one
corresponding to gas confined by the mirror
(Fig.~\ref{Fig:setup}(a) upper figure) and the second one to the
expanding gas (Fig.~\ref{Fig:setup}(a) lower figure). The
entanglement of the spin with a macroscopic number of atoms can be
interpreted as a macroscopic quantum gate, as explained in
Fig.~\ref{Fig:qnd}), implementing a {\em quantum nondemolition
interaction} (QND) \cite{qnd}. In this sense the setup represents
a ``amplifier'' of the state of the qubit. This situation is
reminiscent of a Single Electron Transistor (SET) in mesoscopic
physics \cite{set}, and has stimulated the name {\em Single Atom
Transistor} (SAT) for the setup Fig.~\ref{Fig:setup}(a) in
Ref.~\cite{Micheli}, with the essential difference that the
dynamics underlying (\ref{Eq:MacroCATstate}) is completely
coherent. We finally remark that this setup also allows for a {\em
single shot QND measurement} of the impurity atom (qubit) by
observing in a single experiment the distinct properties of the
$|\phi_\downarrow(t)\rangle$ or $|\phi_\uparrow(t)\rangle$ quantum
phases.

As a variant of the configuration of Fig.~\ref{Fig:setup}(a) we will
consider below in particular the case where the quantum gas is loaded
in an optical lattice, as illustrated in Fig.~\ref{Fig:setup}(b). In
this case the gas could be loaded initially, for example, in a Mott
insulating state, i.e. where large repulsion of the gas atom leads to
a filling of the lattice sites with exactly one atom per lattice site
\cite{Jaksch,Greiner,Stoeferle}. The cat state (\ref{Eq:MacroCATstate})
will thus correspond to a superposition of the {\em Mott phase} and
the {\em melted} Mott phase, i.e. a {\em (quasi-) condensate} of atoms
obtained by expansion of the atomic gas:
\begin{eqnarray}
  |\Psi(t)\rangle =
  \alpha_\downarrow|\downarrow\rangle|\textrm{BEC}\rangle+\alpha_\uparrow|\uparrow\rangle|\textrm{Mott}\rangle.\label{Eq:MacroCATstate2}
\end{eqnarray}
In this case the distinguishing features of the two entangled
quantum phases are the observation / non-observation of
interference fringes as signatures of the Mott and BEC phase, when
the atomic gas in released in a single experiment.

Transport through an impurity is a well studied problem in mesoscopic
condensed matter physics \cite{Datta,Mahan,Levitov,Cazalilla}, which
typically focuses on conductance properties of a system attached to
leads. In contrast, in the context of cold gases we have a full
time-dependent coherent dynamics in an otherwise closed system.

A short summary of the present work including results from
numerical studies was presented in Ref.~\cite{Micheli}. In this
paper we will present details of our analytical calculations,
while we refer to Ref.~\cite{Daley} on a complementary numerical
treatment of these problems using time dependent DMRG techniques.

The paper is organized as follows: In Sec.~\ref{Sec:Model} we
introduce the model used for describing the implementation of the
Single Atom Mirror using cold atoms in optical lattice. In
Sec.~\ref{Sec:Single_Particle_Scattering} we consider the detailed
scattered processes involved in the transport of a single particle
through the mirror. We solve exactly the scattering problem in the
lattice by integrating the Lippmann-Schwinger Equation (LSE) and
discuss the obtained scattering amplitudes and spectrum of the
bound states. Finally, in Sec.~\ref{Sec:Many_body_scattering} we
generalize the discussion to the case of interacting many-systems
including the cases of a 1D degenerate Fermi-gas, a 1D
quasi-condensate and Tonks gas.

\section{Model}\label{Sec:Model}

In this section we introduce the our model system by specifying
the Hamiltonian for a 1D lattice gas coupled to an impurity, and
we explain the key idea behind our setup. We will start with a
discussion of spin-dependent collisions between the gas and the
impurity, and then present the central idea of quantum
interference as a way to switch atomic transport.

\subsection{Effective Spin-Dependent Hamiltonian}

We consider the dynamics of a spin-1/2 atomic impurity $Q$ coupled to
a 1D quantum gas of either bosonic or fermionic probe atoms $A$. The
Hamiltonian for system is split into three parts as
\begin{eqnarray}
  H &=& H_A + H_Q + H_{AQ}\label{Eq:Hamiltian_A_Q_AQ}.
\end{eqnarray}
Here $H_Q$ ($H_A$) describes the uncoupled dynamics of the impurity
atom $Q$ (the degenerate quantum gas of probe atoms $A$), while
$H_{AQ}$ accounts for the interaction between the two atomic species,
$Q$ and $A$.

A degenerate quantum gas of bosonic or fermionic atoms $A$ trapped in
the lowest band of a 1D optical lattice is well described by a Hubbard
model \cite{JakschZollerReview}
\begin{eqnarray}
  H_A &=& \sum_j E_{A,j} a_j^\dag a_j - J\sum_{\langle ij \rangle} a_i^\dag a_j + \frac{U}{2} \sum_j
  a_j^\dag a_j^\dag a_j a_j,\label{Eq:Hamiltian_A}
\end{eqnarray}
where $a_j^\dag$ ($a_j$) are the creation (annihilation) operators
for an atom $A$ on the site $j$, which obey standard commutation
(anticommutation) relations for the case of bosonic (fermionic)
atoms $A$. Moreover, $E_{A,j}$ account for the shift of the bare
energy of an atom localized on the site $j$ in the presence of an
external (e.g. magnetic) shallow trap, $J$ is the tunneling matrix
element for neighboring sites $\langle ij \rangle$ and $U$ gives
the collisional interaction, i.e. the onsite-shift for two atoms
$A$ localized within the same well (which would be zero for the
case of fermions in the same internal state). Denoting the
scattering-length of the atoms $A$ by $a_s$, and their mass by
$m$, we have $U=4\pi\hbar^2a_s\int d^3r |w_j({\bf r})|^4/m$, where
$w_j({\bf  r})$ is the Wannier function for a particle localize on
the site $j$.

In the present setup we regard the impurity atom $Q$ to be trapped
within a tight one-dimensional lattice, as depicted in
Fig.~\ref{Fig:setup}(b). Therefore, we may restrict ourselves to
the lowest trap-state of the $j=0$ well for the internal states
$\sigma=\downarrow,\uparrow$, respectively. The uncoupled dynamics
of the impurity corresponds to spin-1/2 system, i.e.
\begin{eqnarray}
  H_Q &=& \sum_\sigma E_{Q,\sigma} |Q_\sigma\rangle\langle Q_\sigma|,\label{Eq:Hamiltian_Q}
\end{eqnarray}
where $|Q_\sigma\rangle$ ($E_{Q,\sigma}$) denotes the state (energy)
of the atom $Q$ with spin $\sigma=\downarrow,\uparrow$.

Given the tight trapping of the impurity atom, the interaction of
probe and impurity atom is restricted to the site of the impurity,
an in general has the form of an effective spin-dependent
collisional interaction
\begin{eqnarray}
  H_{AQ} &=& W_{{\rm eff},\uparrow}
  |\uparrow\rangle\langle \uparrow| a_0^\dag a_0 + W_{{\rm eff},\downarrow}
  |\downarrow\rangle\langle \downarrow| a_0^\dag a_0,
\end{eqnarray}
where $a_0^\dag$ ($a_0$) is the creation (annihilation) operator for a
probe atom on the site of the impurity, $j=0$. Here, $W_{{\rm
    eff},\sigma}=4\pi\hbar^2 a_{\sigma}/\mu \int d^3r |w_0({\bf r})|^2
|\psi_{Q,\sigma}({\bf r})|$ denotes the effective interaction for
a probe atom $A$ and the impurity atom $Q$ in state $\sigma$ in
terms of their effective scattering length $a_{\sigma}$ and $\mu$
is the reduced mass for $A$ and $Q$. The effective tunneling rate
of a probe atom with energy $E$ through the impurity is then given
by $J_{{\rm eff},\sigma} = J^2/(E-W_{{\rm eff},\sigma})+{\cal
O}(J^4)$ for the qubit in state $\sigma$.  An obvious way to
provide for a spin-dependent single atom mirror is to have the
effective interaction for one spin state as large as possible
($|W_{{\rm eff},\uparrow}|\gg J$), thus blocking the transport
of the probe atoms through the impurity site, while for the other
it is effectively not present, ($|W_{{\rm
    eff},\downarrow}|\ll J$). This can be achieved, for example, by tuning the
internal state dependent scattering length $a_\sigma$ or by
engineering the spin-dependent trapping \cite{JakschZollerReview}.
The quality of the qubit dependent switch then depends on the
difference of the moduli of the effective interactions, $|W_{{\rm
eff},\uparrow}|-|W_{{\rm eff},\downarrow}|$. Thus the goal an
efficient scheme is to make $|W_{{\rm
    eff},\uparrow}|-|W_{{\rm eff},\downarrow}|$ as large as possible
and obtain $|W_{{\rm eff},\uparrow}|\gg J \gg |W_{{\rm
    eff},\downarrow}|$.

\subsection{Controlling the transport by interference}

In this section we will show now that with the help of quantum
interference we can engineer an effectively infinite (zero) atomic
repulsion, $W_{{\rm eff},\uparrow}\rightarrow\infty$ ($W_{{\rm
    eff},\downarrow}\rightarrow0$), for the qubit in state
$\sigma=\downarrow$ ($\sigma=\uparrow$). This is equivalent to tuning
the Feshbach resonance to the point of infinite (zero) scattering
length.

\begin{figure}
  \begin{center}
    \includegraphics[width=0.8\columnwidth]{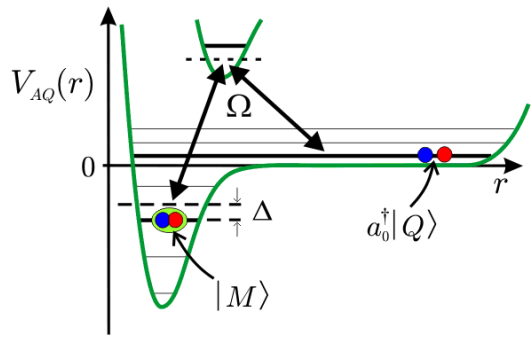}
    \caption{The optical Feshbach setup couples the atomic state
      $a^\dag_0|Q\rangle$ (in a particular motional state quantized by
      the trap) to a molecular bound state $|M\rangle$ of the
      Born-Oppenheimer potential $V_{AQ}(r)$ with effective Rabi
      frequency $\Omega$ and detuning $\Delta$.}\label{Fig:OptFB}
  \end{center}
\end{figure}

The quantum interference mechanism required to engineer the
described spin-dependence of $W_{{\rm eff},\sigma}$ is obtained by
exploiting the properties of either an optical or a magnetic
Feshbach resonance. In the case of an optical Feshbach resonance a
Raman laser drives the transitions from the joint state of the two
atoms on the impurity site, $a_0^\dag|Q_\sigma\rangle\equiv
|Q_\sigma\rangle\otimes a_0^\dag|{\rm vac}\rangle$, via an
off-resonant excited molecular state back to a bound
hetero-nuclear molecular state $|M_\sigma\rangle$ in the lowest
electronic manifold (see Fig.~\ref{Fig:OptFB}). The Raman
processes is described by the effective two-photon Rabi frequency
$\Omega_\sigma$ and detuning $\Delta_\sigma$ for each
spin-component $\sigma$. For the case of a magnetic Feshbach
resonance, the effective Hamiltonian has the same form, but with
$\Omega_\sigma$ being the coupling strength between the open and
closed channels and $\Delta_\sigma$ the detuning of the magnetic
field. The Hamiltonian describing the interaction between the
probe atoms and the impurity is \cite{Holland}
\begin{gather}
  H_{AQ} = \sum_\sigma \Big{[} E_{M,\sigma}
  |M_\sigma\rangle\langle M_\sigma| + \Omega_\sigma
  \left(|M_\sigma\rangle\langle Q_\sigma|a_0 + {\rm
      h.c.}\right)\nonumber\\
    +W_{Q,\sigma}|Q_\sigma\rangle\langle
  Q_\sigma| a_0^\dag a_0 + W_{M,\sigma}|M_\sigma\rangle\langle
  M_\sigma| a_0^\dag a_0 \Big{]},\label{Eq:Hamiltian_AQ}
\end{gather}
where the bare energy of the molecular bound state is
$E_{M,\sigma}=E_{A,0}+E_{Q,\sigma}+\Delta_\sigma$. Here the first two
terms describe the resonant coupling induced by the Feshbach
mechanism, while the last two describe the off-resonant collisions
between an atom $A$ and an atom $Q$ (a molecule $M$) in state $\sigma$
by means of their on-site shift $W_{Q,\sigma}$ ($W_{M,\sigma}$) for
the impurity site. The Hamiltonian~\eqref{Eq:Hamiltian_A_Q_AQ}
conserves the spin-component of the impurity,
$S_\sigma\equiv|Q_\sigma\rangle\langle
Q_\sigma|+|M_\sigma\rangle\langle M_\sigma|$, i.e. $[H,S_\sigma]=0$.
Therefore, we can consider the dynamics for the two spin components of
$Q$ separately, and in the following we will drop the spin index
$\sigma$ and choose the reference energy as $E_{A,0}=E_{Q,\sigma}=0$.

\begin{figure}
  \begin{center}
    \includegraphics[width=.8\columnwidth]{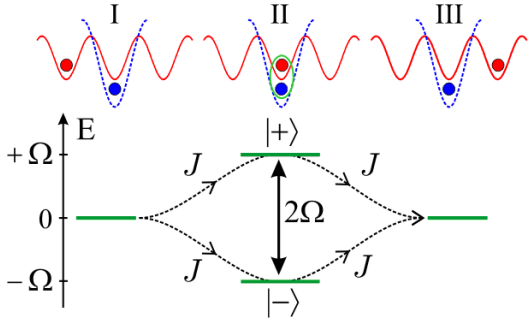}
    \caption{ A single atom
      passes the impurity (I$\rightarrow$III) via the two dressed
      states (II), $|+\rangle=a^\dag_0|Q\rangle+|M\rangle$ and
      $|-\rangle=a^\dag_0|Q\rangle-|M\rangle$ and quantum
      interference gives rise to an effective tunneling rate $J_{\rm
        eff}$.}\label{Fig:EIT}
  \end{center}
\end{figure}

For off-resonant laser driving ($|\Delta| \gg \Omega$), the Feshbach
resonance enhances the interaction between $A$ and $Q$ atoms, giving
the familiar result $W_{\rm eff} = W_{Q}+\Omega^2/\Delta$. However,
for resonant driving ($\Delta=0$) the physical mechanism changes, and
the effective tunneling $J_{\rm eff}$ of an atom $A$ past the
impurity (Fig.~\ref{Fig:EIT}, $\rm I \rightarrow III$) is blocked by
quantum interference. On the impurity site, laser driving mixes the
states $a_0^\dag|Q\rangle$ and $|M\rangle$, forming
two dressed states with energies
$E_\pm=W_{Q}/2 \pm
(W_{Q}^2/4+\Omega^2)^{1/2}$ (Fig.~\ref{Fig:EIT}, II). The two
resulting paths for a particle of energy $E$ destructively interfere
so that for large $\Omega \gg J$ and $W_{Q}=0$,
\begin{eqnarray}
  J_{\rm eff} &=& -\frac{J^2}{E + \Omega}-\frac{J^2}{E -
    \Omega}\rightarrow 0.\nonumber
\end{eqnarray}
This is analogous to the interference effect underlying
Electromagnetically Induced Transparency (EIT) \cite{eit}, and is
equivalent to having an effective interaction $W_{\rm
eff}\rightarrow \infty$. In addition, if we choose
$\Delta=\Omega^2/W_{Q}$, the paths interfere constructively,
screening the background interactions $W_Q$ to produce perfect
transmission ($W_{\rm eff}\rightarrow 0$). The insensitivity of
the interference scheme to losses from the dressed states due to
their large detuning has been argued in Ref.~\cite{Micheli}.

\section{Single particle scattering from an
  impurity}\label{Sec:Single_Particle_Scattering}

In this section we will analyze the scattering of a single probe atom
$A$ from an impurity atom $Q$. We will formulate the scattering
problem, then solve the time-independent and time-dependent
Schr\"{o}dinger Equation, to finally obtain the dynamics of
wave-packets in the lattice.

We consider a probe atom $A$ approaching the impurity from the left,
as a plain Bloch-wave with quasi-momentum $k$. Hence the state of the
system is given by
\begin{equation}
  |k\rangle =\left( \frac{a}{2\pi }\right)
  ^{1/2}\sum_{j}e^{ikx_{j}}|j\rangle, \label{Eq:Psik_2particle}
\end{equation}
where $|j\rangle =|Q\rangle\otimes a_j^\dag|{\rm vac}\rangle$ is the
joint state of the atoms $A$ and $Q$, with $A$ ($Q$) localized in the
lowest vibrational state of the well $j$ (the impurity well $j=0$) and
$a$ is the lattice spacing.

The free evolution of the system is given by the hopping of the atoms
$A$ between neighboring sites at the tunneling rate $J$, whereas the
composite molecule $M$ is detuned by $\Delta$ from the threshold for
the joint state of $A$ and $Q$. Thus, with $E_k=-2J\cos ka$ being the
energy of a Bloch-wave in the first Bloch-band with quasi-momentum
$k$, we have
\begin{eqnarray}
  H_0 &=& -J \sum_j \left( |j+1\rangle\langle j| + \mathrm{h.c.} \right) +
  \Delta | M \rangle\langle M|  \nonumber\\
  &=& \int_{-\pi/a}^{\pi/a} dk E_k |k\rangle\langle k| + \Delta
  |M\rangle\langle M|,\label{Eq:H0_2particle}
\end{eqnarray}
where $|M\rangle$ denotes the molecular bound state localized on the
impurity site. From Eq.~\eqref{Eq:H0_2particle} we obtain the
propagation of the incoming plane wave $|k\rangle$ at group-velocity
$v_k=\partial E_k/\partial k = 2Ja\sin ka$ in the first Bloch band.

Due to the strong confinement of the particles $A$ and $Q$ in the
lattices, their interaction is restricted to the impurity site. There,
their bare interaction induces an on-site-shift $W$ for the joint
atomic state of $A$ and $Q$ on the impurity ($|0\rangle$). Moreover,
the photo-association lasers effectively couple the latter state to
the trapped molecular state ($|M\rangle$) at Rabi-frequency $\Omega$,
yielding
\begin{eqnarray}
  V&=&W|0\rangle\langle0|+\Omega \left(|M\rangle\langle 0|+\mathrm{h.c.}
  \right).\label{Eq:V_2particle}
\end{eqnarray}

\subsection{Scattering solution}

The scattering of a particle $A$ with energy $E=E_{k}$ in the first
Bloch band by the impurity $Q$ is described by a solution of the
Lippmann-Schwinger Equation (LSE). The scattering wave function
$|\phi_+\rangle $ obeys
\begin{equation}
  |\phi_{+}\rangle=|k\rangle +G_{0}(E+i0^{+})V|\phi _{+}\rangle,\label{Eq:LSE}
\end{equation}
with incident plane wave $|k\rangle$ with quasimomentum $k$
($0<k<\pi/a$), and $G_{0}(z)=1/(z-H_{0})$ the free propagator.
Expanding the scattering wave function
\begin{equation}
  |\phi _{+}\rangle=\left(\frac{a}{2\pi}\right)^{1/2} \left[\sum_{j}\alpha
    _{j}|j\rangle+\beta|M\rangle\right]\label{Eq:LSE_ansatz}
\end{equation}
the amplitudes $\alpha_{j}$ and $\beta$ satisfy
\begin{subequations}
  \label{Eq:LSE2}
  \begin{eqnarray}
    \alpha_{j}&=&e^{ikx_{j}}+\mathcal{G}_{j}(E_{k})\left(
      W\alpha_{0}+\Omega\beta\right), \\
    \beta &=&\mathcal{G}_{M}(E_{k})\Omega\alpha_{0}
  \end{eqnarray}
\end{subequations}
with atomic and molecular propagators
\begin{subequations}
  \begin{eqnarray}
    \mathcal{G}_{j}(E) &=&\langle j|G_{0}(E+i0^{+})|0\rangle\nonumber\\
    &\equiv &\frac{a}{2\pi}\int_{-\pi /a}^{+\pi /a}dk\frac{e^{ikx_{j}}}{E-E_{k}+i0^{+}}=\frac{
      e^{ik|x_{j}|}}{iv_{k}/a},\nonumber\\
    \mathcal{G}_{M}(E) &=&\langle M|G_{0}(E+i0^{+})|M\rangle \equiv \frac{1}{
      E-\Delta +i0^{+}}.\nonumber
  \end{eqnarray}
\end{subequations}

Solving Eqs.\eqref{Eq:LSE2} we find
\begin{subequations}
  \begin{eqnarray}
    \alpha _{j} &=&e^{ikx_{j}}+\frac{W_{k}}{iv_{k}/a-W_{k}}e^{ik|x_{j}|},
    \label{Eq:LSE_ampl_atomic} \\
    \beta  &=&\frac{-i\Omega v_{k}/a}{\Omega ^{2}+\left( E_{k}-\Delta
        +i0^{+}\right) \left( W-iv_{k}/a\right) }
  \end{eqnarray}
\end{subequations}
with effective energy dependent interaction
\begin{equation}
  W_{k}=W+\frac{\Omega ^{2}}{E_{k}-\Delta +i0^{+}},
\end{equation}
where we read off the transmission and reflection amplitudes
\begin{subequations}
  \begin{eqnarray}
    t_k &=&\frac{1}{1+iaW_{k}/v_{k}}\\
    r_k &=&\frac{-1}{1-iv_{k}/aW_{k}},
  \end{eqnarray}\label{Eq:tk_rk}
\end{subequations}
respectively.

Note that the presence of the molecular state introduces an effective
energy-dependent interaction $W_{k}$. This can be interpreted in terms
of an effective atomic scattering length with background scattering
length proportional to $W$ and a resonant term, corresponding to an
optical Feshbach resonance at energy given by the detuning from the
molecular state $\Delta$, and width determined by the Rabi frequency $\Omega$.

The scattering matrix
\begin{equation}
  S(E_{k})\mathbb{=}\left(
    \begin{array}{cc}
      r_k & t_k \\
      t_k & r_k%
    \end{array}%
  \right)
\end{equation}%
is unitary, as follows readily from the above expressions \eqref
{Eq:tk_rk}. This implies $T_{k}+R_{k}\equiv
|t_{k}|^{2}+|r_{k}|^{2}=1$. We can assign phase shifts $t_{k}\pm
r_{k}=\exp (i\delta _{k}^{\pm })$ for the symmetric and antisymmetric
states $|k\rangle \pm |-k\rangle $, $\delta _{k}^{+}=-2\arctan
(aW_{k}/v_{k})$ and $\delta _{k}^{-}=0$, respectively, so that
$R_{k}=\sin ^{2}(\delta _{k}^{+}/2)$.

\subsection{Discussion of the Scattering Amplitudes}

\begin{figure}[tb]
  \begin{center}
    \includegraphics[width=0.95\columnwidth]{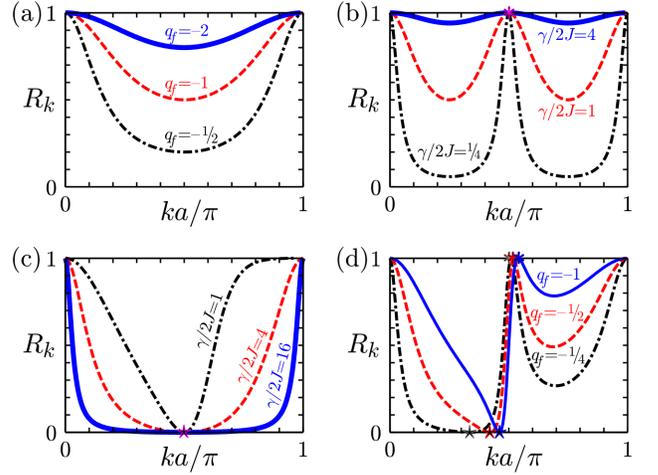}
\end{center}
\caption{Reflection coefficient $R_k$, showing in (a) partial reflection for
  bare background collisions $W\neq0$ but $\Omega=0$, i.e.
  $q_f=-1/2,-1,-2$ (dotted, dashed, full line) and $\gamma=E_R=0$, (b)
  complete reflection at $k_0=\pi/2a$ for $\gamma/2J=1/4,1,4$ (dotted,
  dashed, full line) and $E_R=q_f=0$, cf. $R\approx 1$ for
  $\gamma/2J=4$ (full line), (c) perfect transmission at $k_0=\pi/2a$
  for $\gamma/2J=1,4,16$ (dotted, dashed, full line), $q_f=-2$ and
  $E_R=\gamma/4$, cf. $R\approx 0$ for $\gamma/2J=16$ (full line), (d)
  the asymmetry of the Fano-profiles for $q_f=-1/4,-1,-4$ (dotted,
  dashed, full line), $\gamma/2J=1/2$ and $E_R=0$.}
\label{Fig:Reflection}
\end{figure}

In the absence of molecular couplings ($\Omega =0$) the on-site
interaction $W$ between the B atom and the impurity Q always gives
rise to \emph{partial} reflection and transmission, see
Fig.~\ref{Fig:Reflection}a,
\begin{eqnarray}
R_{k}=1-T_{k}=\frac{W^{2}}{W^{2}+4J^{2}\sin^2\left(ka\right)}<1.\label{Eq:Rk_for_W}
\end{eqnarray}

The significant new feature introduced by the optical Feshbach
resonance is that we can achieve essentially \emph{complete blocking}
($R_{k}=1$) and \emph{complete transmission} ($T_{k}=1$). We obtain
this in the limits $ \Omega \gg J$ and $\Delta =0$, and $\Omega \gg J$
and $\Delta =-\Omega ^{2}/W $, respectively. Physically, the first
case corresponds to tuning to the point of ``infinite" scattering
length, while the second case corresponds to tuning to the point of
``zero'' scattering length, respectively.

In the general case the energy dependence of transmission and
reflection has the form of a Fano-like profile (see
Fig.~\ref{Fig:Reflection}).
In the region $|E_k|\ll 2J$ we may neglect the dispersion effects,
i.e. $v_k \approx 2Ja$, and obtain Fano-line-shapes for transmission
and reflection as
\begin{subequations}
\begin{eqnarray}
T(\varepsilon)&=&\frac{1}{1+q_f^2}\frac{\left(\varepsilon+q_f\right)^2}{\varepsilon^2+1},\\
R(\varepsilon)&=&\frac{1}{1+1/q_f^2}\frac{\left(\varepsilon-1/q_f\right)^2}{\varepsilon^2+1},
\end{eqnarray}\label{Eq:Fano_profile}
\end{subequations}
where $\varepsilon\equiv (E-E_R)/(\gamma/2)$ is the dimensionless
energy in units of the resonance width $\gamma$, $E_R$ is the
resonance energy and $q_f$ is the Fano-$q$-parameter. These parameters
of the Fano-profile are related to $J,W,\Delta,\Omega$ by
\begin{subequations}
\begin{align}
  \gamma &= \frac{4J\Omega^2}{W^2+4J^2}, \quad & \Omega^2 &= J\gamma\left(q_f^2+1\right),\\
  E_R &= \Delta-\frac{W\Omega^2}{W^2+4J^2}, \quad &\Delta &= E_R - \frac{q_f\gamma}{2},\\
  q_f &= \frac{-W}{2J}, \quad & W &= -2J q_f.
\end{align}
\label{Eq:Fano_parameter}
\end{subequations}
For $W=0$ the asymmetry parameter $q_f$ vanishes, the reflection
profile is symmetric, and for $\Omega>0$ resembles a
Breit-Wigner-profile, see Fig~\ref{Fig:Reflection}(b). The maximum
$R_k=1$ is attained at $\varepsilon=-q_f$ ($E_k=\Delta$,
$|\Delta|<2J$) and has a width $\gamma=\Omega^2/J$.

For finite background collisions $q_f\neq 0$ ($W\neq0$) the
transmission profile is asymmetric, and shows an additional minimum
$R_k=0$ at $\varepsilon=-1/q_f$ ($E_k=\Delta-\Omega^2/W=\Delta_\star$,
$|\Delta_\star|<2J$), see Fig~\ref{Fig:Reflection}(c).

Near the edges of the Bloch band, $k_\pm=(\pi\mp\pi)/2a$
($E_k=\pm2J$), transmission and reflection deviate from the Fano
line-shape Eq.~\eqref{Eq:Fano_profile}. There the group-velocity
$v_k\rightarrow 0$ and thus also the transmission $T_k$ vanishes,
unless the dressed resonance $\Delta_\star$ is tuned to respective
edge of the Bloch band. The transmission coefficient are given by
\begin{subequations}
  \begin{eqnarray}
    T_{k{\approx}k_\pm}\approx
    \frac{4J^2a^2\left(k-k_\pm\right)^2}{\left[W-\Omega^2/\left(\Delta\mp2J\right)\right]^2}
    &\mbox{for}& \Delta_\star\neq \pm 2J,\nonumber\\
    T_{k{\approx}k_\pm}\approx 1-\frac{W^4 a^2\left(k-k_\pm\right)^2}{2\Omega^4}
    &\mbox{for}& \Delta_\star = \pm 2J.\nonumber
  \end{eqnarray}
\end{subequations}

The reflection coefficient $R_{k}$ as a function of energy
$-2J<E_{k}<+2J$ is shown in Fig.~\ref{Fig:Reflection}.

In the absence of molecular coupling, $\gamma=0$ ($\Omega=0$), the
reflection is unity at the band-edges, $k=0,\pi/a$, and decreases
within the Bloch-band due to the increase of the group-velocity
(see Fig.\ref{Fig:Reflection}(a)). The profile is symmetric about
the middle of the Bloch band, $k=\pi/2a$, where it attains its
\emph{minimum},
\begin{eqnarray}
  R_k=\frac{1}{1+q_f^{-2}}=\frac{1}{1+\left(2J/W\right)^2}.\nonumber
\end{eqnarray}

In the presence of molecular couplings, $\gamma\neq 0$ ($\Omega>0$), and for
$E_R=q_f=0$ the reflection profile is still symmetric about
$k=\pi/2a$  (see
Fig.\ref{Fig:Reflection}(b)). However, now it approaches its \emph{maximum}, $R_k=1$, at $k=\pi/2a$,
and has now two minima at $k\approx \pi/4a$ and $k\approx 3\pi/4a$,
given by
\begin{eqnarray}
  R_k=\frac{1}{1+\left(2J/\gamma\right)^2}=\frac{1}{1+\left(\sqrt{2}J/\Omega\right)^4}.\nonumber
\end{eqnarray}

For $\gamma,q_f\neq 0$ we obtain an asymmetric Fano-profile
(see Fig.\ref{Fig:Reflection}(d)), which for
$|E_R+q_f\gamma/2|=|\Delta|<2J$ shows complete reflection, $R_k=1$, at
$\varepsilon=-q_f$, while for $|E_R-\gamma/2q_f|<2J$ one has
perfect transmission, $R_k=0$, at $\varepsilon=+1/q_f$.

The reflective and transmissive resonance are present regardless
of the magnitude of $q_f$, and their width is $\propto\gamma$.
However, for $\gamma\leq 8J |q_f+1/q_f|$ they may both occur
within the physical energy range of the Bloch-band (see
Fig.\ref{Fig:Reflection}(d)), while for $\gamma > 8J |q_f+1/q_f|$
only one resonance appears (see Fig.\ref{Fig:Reflection}(b,c)).
Thus in the limit $\gamma \gg 8J |q_f+1/q_f|$ we achieve complete
blocking , $R_k=1$ for all $k$, by tuning $E_R\approx-q_f\gamma/2$
(see full-line in Fig.\ref{Fig:Reflection}(b)). Within the same
limit we can also efficiently screen any background-interaction
$W$ and achieve complete transparency, $T_k=1$ for all $k$, by
tuning $E_R\approx\gamma/2q_f$ (see full-line in
Fig.~\ref{Fig:Reflection}(c)).

\subsection{Interference mechanism}

Physically, the features of complete blocking ($R_k=0$) and complete
transmission ($T_k=0$) are induced by an interference mechanism, as
the probe atom may tunnel via two interfering paths of dressed atomic
+ molecular states, as depicted in Fig.~\ref{Fig:EIT}.

For simplicity we start by elucidating the underlying interference
mechanism (present for $\Omega\neq0$) in the regime of strong
coupling, $\Omega^2\gg (2J)^2+|W\Delta|$. In this regime we can
consider the local dynamics within the individual sites, and treat the
tunneling $J$ by means of perturbation theory.

To zeroth order in $J$ the Hamiltonian $H$ decouples the dynamics of
the individual sites $j$ as
\begin{eqnarray}
H^{(0)} =&
W|0\rangle\langle0|&+\Omega|0\rangle\langle
{M}|+\nonumber\\
+&\Omega|M\rangle\langle
0|&+\Delta|M\rangle\langle{M}|.\label{Eq:Ham_J_zero}
\end{eqnarray}
Outside the impurity ($j\neq 0$) its eigenstates are the joint states
of the atoms $A$ and $Q$, $|j\rangle$, with energy $E_0=0$, whereas on the
impurity ($j=0$) the strong coupling $\Omega$ between the atomic state
$|0\rangle$ and the molecular state $|M\rangle$ induces the two states
to split into two dressed state $|E_\pm\rangle$ of atoms + molecules
with energy $E_\pm$, see Fig.~\ref{Fig:EIT}. By diagonalizing the
$2\times2$-matrix in Eq.~\eqref{Eq:Ham_J_zero} we obtain the
amplitudes and energy of the dressed states as
\begin{subequations}
\begin{eqnarray}
|E_\pm\rangle &=& \left[\frac{1\pm\xi}{2}\right]^{1/2}|0\rangle \pm
\left[\frac{1\mp\xi}{2}\right]^{1/2}|M\rangle,\label{Eq:Dressed_States_Local}\\
E_\pm &=&
\frac{W+\Delta}{2}\pm \left[\Omega^2+\left(\frac{W-\Delta}{2}\right)^2\right]^{1/2},\label{Eq:Dressed_Energy_Local}\\
\xi &=& \frac{W-\Delta}{\left[\left(W-\Delta\right)^2+4\Omega^2\right]^{1/2}},
\end{eqnarray}
\end{subequations}
where $\xi$ characterizes the asymmetry of the amplitudes, i.e. for
$\xi=0$ ($W=\Delta$) the dressing is completely symmetric while for
$|\xi|=1$ ($\Omega=0$) the atomic and molecular state decouple.
From Eq.~\eqref{Eq:Dressed_Energy_Local} we see that for
$\Omega\gg|W\Delta|$ the dressed states are far off-resonant from
$E=0$, and hence will be only virtually populated.

The effects of the hopping of the atom $A$ on the $E\approx0$ modes
$|j\neq 0\rangle$ can be accounted by means of an effective
Hamiltonian $H_{\rm eff}$. Following Ref.~\cite{Cohen} we obtain the
dynamics as a perturbative series in the hopping amplitude $J$,
$H_{\rm eff}=H^{(0)}+H^{(1)}+H^{(2)}+\ldots$.  To first order in $J$
one obtains
\begin{eqnarray}
H^{(1)}&=& -J \sum_{j<0}|j\rangle\langle{j-1}|
-J\sum_{j>0}|j\rangle\langle{j+1}|+{\rm h.c.} \nonumber\\
&=& \sum_{\alpha=L,R}\int dk E_k |k_\alpha\rangle\langle k_\alpha|,
\end{eqnarray}
where $|k_L\rangle$ ($|k_R\rangle$) are the Bloch-waves with quasi-momentum $k$
on the left (right) side of the impurity site,
\begin{eqnarray}
|{k_{L,R}}\rangle &=& \left(\frac{a}{2\pi}\right)^{1/2}\sum_{\pm j>0}
e^{+ikx_j} |j\rangle.\nonumber
\end{eqnarray}
This the flat dispersion relation $E^{(0)}=0$ on the left and
right side of the impurity is bent to
$\varepsilon(k)=-2J\cos(ka)$, i.e. we recover the Bloch-band(s).

To second order in $J$ we obtain
\begin{gather}
  H^{(2)} = -J_{\rm eff}
  \sum_{i,j=\pm1}|i\rangle\langle j|,\\
  J_{\rm eff} = \frac{J^2}{2}\left[\frac{1+\xi}{E_0-E_+} + \frac{1-\xi}{E_0-E_-}\right]
  = \frac{J^2\Delta}{\Omega^2-W\Delta}.\label{Eq:Jeff_PT2}
\end{gather}
We see that tuning on resonance $\Delta=0$ the two contributions
in Eq.~\eqref{Eq:Jeff_PT2} cancel each other as
\begin{eqnarray}
  J_{\rm eff} = \frac{J^2}{\sqrt{W^2+\Omega^2}} -
  \frac{J^2}{\sqrt{W^2+\Omega^2}}=0,\nonumber\\
\end{eqnarray}
which gives perfect blocking by the impurity.
Furthermore, from
Eq.~\eqref{Eq:Dressed_Energy_Local} we obtain that for
$\Delta_{\star}=\Delta-\Omega^2/W=0$ one of the dressed states
$|E_B\rangle$ becomes a resonance for an incoming particle
($E_B=E_0=0$) and for $\Omega\neq 0$ provides for complete transmission by
means of photo-assisted tunneling.  The described interference
mechanism induced by the optical Feshbach resonance is in marked
contrast to the situation where one has background collisions.
There the particle $A$ can tunnel only via one path through the impurity
($|\xi|=1$), and therefore the effective hopping
rate is always finite, i.e. $J_{\rm eff}=-J^2/W\neq 0$.

\begin{figure}[tb]
  \begin{center}
    \includegraphics[width=0.95\columnwidth]{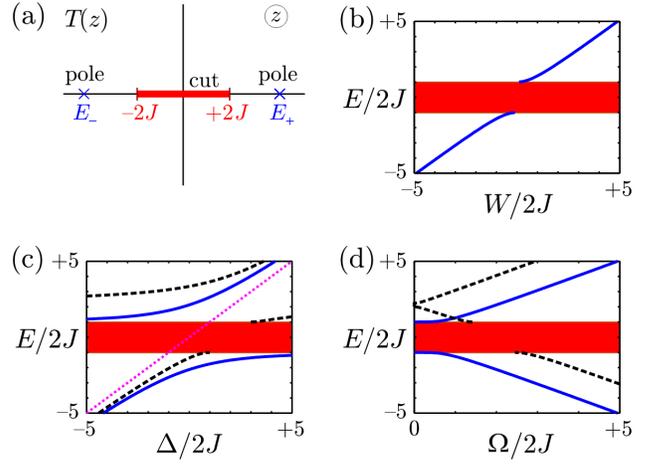}
  \end{center}
  \caption{(a) The analytic properties of the $T$-matrix
    in the complex plane. The cut on the real axis is due to the
    propagation of (dressed) Bloch-waves. The two poles on the
    real axis correspond to two bound states and give rise to
    interference. (b-d) The spectrum $E_n$ of the Hamiltonian $H=H_0+V$ as a
    function of (a) the on-site-shift $W$ on the impurity for
    $\Omega=0$, (b) the detuning $\Delta$ for $\Omega/J=2$ and
    $W=0$ (solid lines), for $\Omega/J=4$ and $W=0$ (dashed lines),
    for $\Omega/J=2$ and $W/J=4$ (dash-dotted lines), (c) the
    Rabi-frequency $\Omega$ for $\Delta=W=0$ (solid lines), for
    $\Delta/J=4$ and $W=0$ (dashed lines), for $\Delta/J=W/J=4$
    (dash-dotted lines).  The continuous (scattering) spectrum, i.e.
    the Bloch-band $E_k$, is indicated as a shaded region
    $-2J<E_k<2J$, whereas the bound-state(s) $E_B$ are shown as
    (solid, dashed, dash-dotted) lines.}\label{Fig:Boundstates}
\end{figure}

\subsection{Discussion of Bound-states}

For completeness we here derive the exact bound-state spectrum of $H$.
For the exact scattering solution, detailed in
Sec.~\ref{Sec:Single_Particle_Scattering}, the bound-states take the
role of dressed states $|E_\pm\rangle$, which are responsible for
the interference mechanism. We will show that for arbitrary $\Omega$
there are always \emph{two} bound-states, provided
$|\Delta-\Omega^2/W|>2J$. For
$|\Delta-\Omega^2/W| \leq 2J$ one of the bound-states turns into a
resonance, which makes the impurity completely transparent for the
atom $A$, $T_k=1$.
Furthermore, we will show that the bound-states for
extends over several lattice sites for $\Omega,W,\Delta \sim J$.
This is in marked contrast to the perturbative result, where the
dressed states were localized on the impurity site, cf. Eq.~\eqref{Eq:Dressed_States_Local}.

We obtain the bound states wavefunctions $|\phi_B\rangle$ from the
homogeneous Lippmann-Schwinger equation
\begin{eqnarray}
|\phi_B\rangle &=& G_0(E)V|\phi_B\rangle,\label{Eq:LSE_homogen}
\end{eqnarray}
where $\phi_B$ denotes the bound-state with energy $E\equiv E_B$
($|E_B|>2J$). Using the ansatz
\begin{eqnarray}
|\phi_B\rangle &=& \sum_j \alpha_j |j\rangle + \beta|M\rangle\label{Eq:Single_Ansatz}
\end{eqnarray}
we find that the atomic and molecular amplitudes, $\alpha_j$ and
$\beta$, satisfy
\begin{subequations}
\begin{eqnarray}
a_j &=& {\cal G}_j(E_B)\left[W\alpha_0+\Omega\beta\right],\\
\beta &=& {\cal G}_M(E_B)\Omega \alpha_0.
\end{eqnarray}
\label{Eq:LSE2_homogen}
\end{subequations}
The atomic and molecular propagators are given by
\begin{subequations}
\begin{eqnarray}
{\cal G}_j(E_B) &=& \langle j | G_0(E_B) | 0 \rangle =
\frac{e^{-|x_j|/r_B}\left[{\rm sign}\left(-E_B\right)\right]^j}{-2J\sinh\left(\kappa_B a\right)},\nonumber\\
{\cal G}_M(E_B) &=& \langle M | G_0(E_B) | 0 \rangle = \frac{1}{E_B-\Delta},\nonumber
\end{eqnarray}
\end{subequations}
and $r_B=a/{\rm acosh}(|E_B|/2J)>0$ denotes the size of the bound-state.

For convenience we first consider the case $\Omega=0$. There the
molecular state decouples from the atomic ones, and we have one
bound-state $|\phi_1\rangle=|M\rangle$ with energy $E_1=\Delta$.
Moreover, for $W\neq0$ we have another bound-state
$|\phi_2\rangle$ with energy $E_2={\rm sign}(W)\sqrt{W^2+(2J)^2}$.
Its amplitudes are given by $\beta=0$ and
\begin{eqnarray}
\alpha_j=\left[{\rm tanh}\left(\frac{a}{r_B}\right)\right]^{1/2}e^{-|x_j|/r_B}\left[{\rm
    sign}(-W)\right]^{j},\nonumber
\end{eqnarray}
with the size $r_B=a/{\rm arccos}\sqrt{1+(W/2J)^2}$. The spectrum
of the system is plotted in Fig.~\ref{Fig:Boundstates}(b) as a
function $W$. For attractive (repulsive) interaction $W<0$ ($W>0$)
the energy of the bound-state $\phi_2$, lies below (above) the
Bloch-band, i.e. $E_2<-2J$ ($E_2>2J$), respectively. For $|W|<2J$
bound state $\phi_2$ extends over several lattice sites, while for
$|W|\gg 2J$ it is localized on the impurity.

In the case $\Omega\neq 0$ a nontrivial solution of Eq.~\eqref{Eq:LSE2_homogen} requires
\begin{eqnarray}
-E_B\left[1-\left(\frac{2J}{E_B}\right)^2\right]^{1/2} &=& W + \frac{\Omega^2}{E_B-\Delta},\label{Eq:Energy_Boundstate}
\end{eqnarray}
which determines the bound-state spectrum $E_B(W,\Delta,\Omega)$.
From Eq.~\eqref{Eq:LSE2_homogen} we obtain the atomic and molecular amplitudes as
\begin{subequations}
\begin{eqnarray}
\alpha_j &=& \beta
\frac{E_B-\Delta}{\Omega}e^{-|x_j|/r_B}\left[{\rm sign}\left(-E_B\right)\right]^j,\label{Eq:Bound_state_atomic_amplitude}\\
\beta &=& \left[1+\frac{\left(E_B-\Delta\right)^2}{\Omega^2\sqrt{1-\frac{4J^2}{E_B^2}}}\right]^{-1/2}.
\end{eqnarray}
\end{subequations}
We solve Eq.~\eqref{Eq:Energy_Boundstate} by expressing
one of the parameters, either $W$, $\Omega$ or $\Delta$, in terms of the
bound-state energy $E_B$,
\begin{subequations}
\begin{eqnarray}
\Delta\left(E_B\right) &=&
E_B+\frac{\Omega^2}{E_B\sqrt{1-4J^2/E_B^2}+W},\label{Eq:Energy_Boundstate_Reverse_Delta}\\
\Omega\left(E_B\right) &=&
\left[\left(\Delta-E_B\right)\left(E_B\sqrt{1-\frac{4J^2}{E_B^2}}
    +W\right)\right]^{1/2},\label{Eq:Energy_Boundstate_Reverse_Omega}\\
W\left(E_B\right) &=&
E_B\sqrt{1+\frac{4J^2}{E_B^2}}+\frac{\Omega^2}{E_B-\Delta}\label{Eq:Energy_Boundstate_Reverse_W}.
\end{eqnarray}\label{Eq:Energy_Boundstate_Reverse}
\end{subequations}
Inverting the functions Eq.~\eqref{Eq:Energy_Boundstate_Reverse}
for $E_B$ yields $E_B$ as a function of $\Delta$, $\Omega$ and
$W$, respectively. For fixed $\Delta$, $\Omega$, $W$ we carry out
the inversion by plotting in Fig.~\ref{Fig:Boundstates} the r.h.s.
of Eq.~\eqref{Eq:Energy_Boundstate_Reverse}(a,b,c) as a function
of $E_B$. In particular in Fig.~\ref{Fig:Boundstates}(b) we plot
detuning $\Delta$ as a function of $E_B$ for constant $\Omega$ and
$W$, in Fig.~\ref{Fig:Boundstates}(c) we plot Rabi-frequency
$Omega$ as a function of $E_B$ for constant $\Delta$ and $W$, and
in Fig.~\ref{Fig:Boundstates}(d) we plot on-site shift $W$ as a
function of $E_B$ for constant $\Delta$ and $\Omega$. In the
following we will give a detailed discussion of
Fig.~\ref{Fig:Boundstates}.

For no background collisions, $W=0$, and arbitrary detuning $\Delta$,
one always has two bound-states $\phi_{1,2}$ with energy $E_1<-2J$ and
$E_2>2J$, respectively, see solid line in
Fig.~\ref{Fig:Boundstates}(c) corresponding to $\Omega=4J$. For
$|\Delta|\gg\Omega$ the energy one bound-state approaches $\Delta$ and
is wavefunction becomes localized on the impurity, while the energy
the other approaches the Bloch-band and its wavefunction extends over
several lattice-sites.  For $\Delta=0$ the two bound-states are split
symmetrically, and their energies are given by (see solid line in
Fig.~\ref{Fig:Boundstates}(d))
\begin{eqnarray}
E_{1,2}=\pm\sqrt{2J^2+\sqrt{4J^2+\Omega^2}}.
\end{eqnarray}
The symmetric splitting allows for complete
reflection at $E_k=0$. In the limit $\Omega\gg 2J$ we recover the
perturbative result Eq.~\eqref{Eq:Dressed_Energy_Local}, as the energies
of the bound states
 approach $E_\pm = \pm \Omega$, with their wavefunctions given by the
dressed states $|E_\pm\rangle$, cf.~Eq.~\eqref{Eq:Dressed_States_Local}.

For finite onsite shift, $W\neq 0$, we also have two bound-states
provided $|\Delta_\star|\equiv|\Delta-\Omega^2/W|>2J$, see dashed
lines in Fig.~\ref{Fig:Boundstates}(c,d). With increasing detuning
$\Delta$ the energy of the bound-state $\phi_1$ approaches the
Bloch-band from below until crossing it for
$-2J+\Omega^2/W<\Delta<+2J+\Omega^2/W$. In this parameter regime
there is merely one bound state, while the other develops a
resonance. This allows for perfect transparency at
$E_k=\Delta-\Omega^2/W$.

Finally we remark that the size $r_B$ of the bound-states is inversely
proportional to the separation of their energy $E_B$ from their
Bloch-band. Thus for $|E_B|\ll 2J$ the wave-function of the bound-states extends over
several lattice sites. For $|E_B|\gg 2J$ the bound-states are localized on
the impurity and we recover the results of the previous section.

\subsection{Wave-packet dynamics}

As an illustration of the time-dependence of the interference mechanism we simulate the
evolution of a gaussian wave-packet $\psi(t)$ with mean quasi-momentum $k=\pi/2a$
incident from the left of the impurity.

These wave-packets are obtained as superposition of the scattering
solutions $\phi_+$ of Sec.~\ref{Sec:Single_Particle_Scattering}, i.e. their
atomic and molecular amplitudes, $\alpha_j(t)=\langle
j|\psi(t)\rangle$ and $\beta(t)=\langle M|\psi(t)\rangle$, are
obtained as
\begin{subequations}
\begin{eqnarray}
\alpha_{j}(t) &=& \sum_{j'} U_{j,j'}(t) \alpha_{j'}(0),\\
\beta(t) &=& \sum_{j'} U_{M,j'}(t) \alpha_{j'}(0),
\end{eqnarray}\label{Eq:Psi_Propagator_Convolution}
\end{subequations}
with the full propagator for the system given by
\begin{eqnarray}
&&U_{j^{\prime},j}(t) = \sum_B
\frac{e^{-(|x_{j'}|+|x_j|)/r_B-iE_B
    t}{\rm sign}(-E_B)^{j'+j}}{\frac{|E_B|}{\sqrt{E_B^2+4J^2}}
+\left[\frac{\Omega}{E_B-\Delta}\right]^2}+
\nonumber\\
&&+\frac{a}{2\pi}\int dk e^{-iE_k t}\left[ e^{+ik|x_{j'}-x_j|}+
  r_k e^{+ik(|x_{j'}|+|x_j|)}\right],\nonumber\\
&&U_{M,j}(t) = \sum_B
\frac{e^{-|x_j|/r_B-iE_B
    t}{\rm sign}(-E_B)^j}{\sqrt{1+\frac{|E_B|(E_B-\Delta)^2}{\Omega^2\sqrt{E_B^2-4J^2}}}}+
\nonumber\\
&&+\frac{a}{2\pi}\int dk e^{-iE_k t}\left[ \frac{\Omega
    t_k}{E_k-\Delta+i0^+} e^{+ik(|x_j|)}\right].
\label{Eq:Propagator_Full}
\end{eqnarray}

\begin{figure}[tb]
  \begin{center}
    \includegraphics[width=0.95\columnwidth]{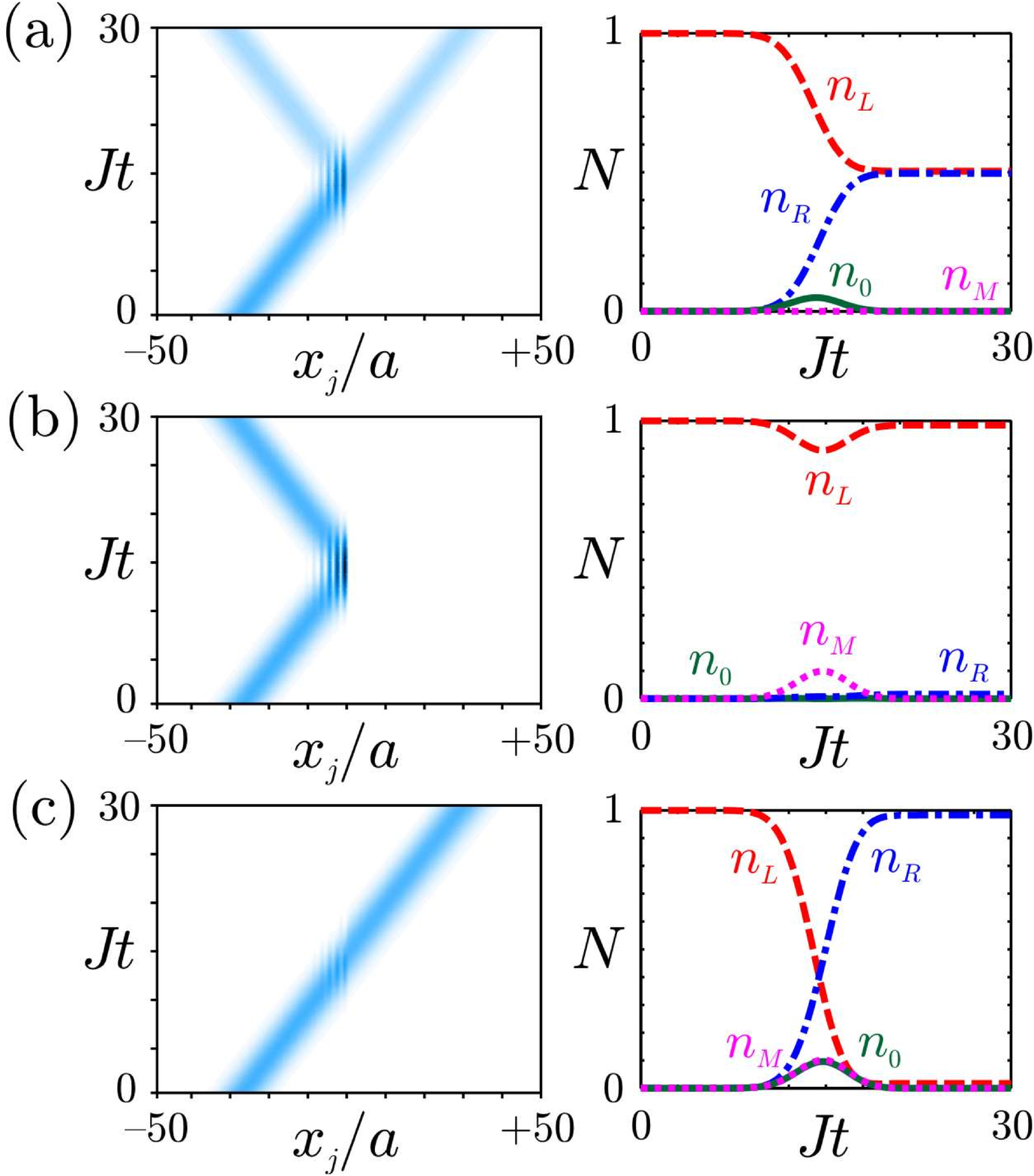}
  \end{center}
  \caption{Evolution of a Gaussian wavepacket with mean quasi-momentum
    $k=\pi/2a$ for $W=2J$ and (a) $\Omega=0$, (b) $\Omega=2J$,
    $\Delta=0$, and (c) $\Omega=\Delta=2J$. On the left side the
    atomic density, $n(x_j,t)$, of atoms $A$ is plotted (darker region
    correspond to higher density).  The right side shows the
    corresponding population of atoms $A$ of the sites to the left,
    $n_L(t)$ (dashed-line), to the right, $n_R(t)$ (dashed-dotted line), and on the
    impurity site $n_0(t)$ (solid line), as well as the population of the molecular
    state, $n_M(t)$ (dotted line).}
 \label{Fig:WavePackets}
\end{figure}

On the left side of Fig.~\ref{Fig:WavePackets} we plot the atomic
populations of the individual sites, $n(x_j,t)=|\alpha_j(t)|^2$. The
right side shows the corresponding atomic populations of the atom $A$
on the left, $n_L=\sum_{j<0}n(x_j,t)$ (dashed line), and on the right
side of the impurity, $n_R=\sum_{j>0}n(x_j,t)$ (dashed-dotted line).
We also plot the population on the impurity, i.e. the atomic
population, $n_0=n(x_0,t)$ (solid line), and the molecular population,
$n_M(t)=|\langle M|\psi(t)\rangle|^2$ (dotted line).  The three
different sets in Fig.~\ref{Fig:WavePackets} correspond to different
coupling strengths, $\Omega$, and detunings, $\Delta$. For all cases
we choose $W=2J$. In Fig.~\ref{Fig:WavePackets}(a) we have $\Omega=0$:
the atom is partially reflected from the impurity with $R_k=1$. In
Fig.~\ref{Fig:WavePackets}(b) we set $\Omega=2J$ and $\Delta=0$, which
gives rise to complete reflection of the wavepacket, $R_k=1$. In
Fig.~\ref{Fig:WavePackets}(c) we have $\Omega=2J$, but now
$\Delta=2J$. We have complete transmission of the atom through the
impurity, $T_k=1$. All this is consistent with the results of
Sec.~\ref{Sec:Single_Particle_Scattering}.

\section{Many body scattering from an impurity}\label{Sec:Many_body_scattering}

In this section we will analyze the evolution of a 1D lattice gas
of \emph{many} atoms $A$ interacting with an impurity atom $Q$.
Since the statistics of the atoms $A$ plays a dominant role, we
will consider the cases of fermionic and bosonic atoms,
separately. In this context we will study analytically the
limiting cases of an ideal Fermi-gas, an ideal Bose-gas and a
Tonks-gas. An exact numerical treatment of the dynamics for the
lattice-gas having arbitrary interaction $U$ is given in
Ref.~\cite{Daley}.

\subsection{Ideal Fermi-gas}\label{Sec:Ideal_Fermi_gas}

We first consider the case, where the probe atoms $A$ are
spin-polarized fermions. The Hamiltonian for the system is given by
\begin{eqnarray}
H &=& -J\sum_j\left(a_j^{\dag}a_{j+1}+a_{j+1}^{\dag}a_{j}\right)+\Delta |M\rangle\langle M|+\nonumber\\
&& +\Omega\left(|M\rangle\langle Q|a_0+a_0^\dag|Q\rangle\langle M|\right)+\nonumber\\
&& + W_Q |Q\rangle\langle Q| a_0^\dag a_0  + W_M |M\rangle\langle{M}|a_0^\dag a_0,\label{Eq:Hamiltonian_Fermions_Basis}
\end{eqnarray}
where the operators $a_j^\dag$ ($a_j$) create (annihilate) an atom $A$ on
site $j$, and obey the canonical anti-commutation relations $\{a_i,a_j^\dag\}=\delta_{ij}$
and $\{a_i,a_j\}=\{a_i^\dag,a_j^\dag\}=0$. Moreover, $|Q\rangle$
($|M\rangle$) denote the states with an atom $Q$ (a molecule in state
$M$) on the impurity, and $W_Q$ ($W_M$) is the onsite-shift for an
atom $A$ and an atom $Q$ (a molecule $M$) on the impurity.

For simplicity henceforth we will restrict ourselves to the case of
equal on-site shifts $W_M=W_Q\equiv W$. In this case we may rewrite the Hamiltonian as
\begin{eqnarray}
H &=& -J\sum_j\left(a_j^{\dag}a_{j+1}+a_{j+1}^{\dag}a_{j}\right)+Wa_0^{\dag}a_0+\nonumber\\
&& \Delta f^\dag f+\Omega\left(f^\dag a_0+a_0^\dag f\right),\label{Eq:Fermi_Spin_Model}
\end{eqnarray}
where the ladder operators $f^\dag\equiv |M\rangle\langle Q|$ and $f\equiv |Q\rangle\langle M|$
obey standard fermionic anti-commutation relations and anti-commute
with $a_j$ and $a_j^\dag$.
The corresponding equations of motions for $a_j$ and $f$ are
linear, provided $W_M=W_Q$. Thus for a Fermi-gas of $N$ atoms $A$ the scattering off the
impurity atom $Q$ will occur independently for each fermion $a_k$ with
scattering amplitudes $t_k$ and $r_k$, according to their
quasi-momentum $k$, cf.~Eq.~\eqref{Eq:tk_rk}. The details of this
calculation will be given below.

\begin{figure}[thb]
  \begin{center}
    \includegraphics[width=0.95\columnwidth]{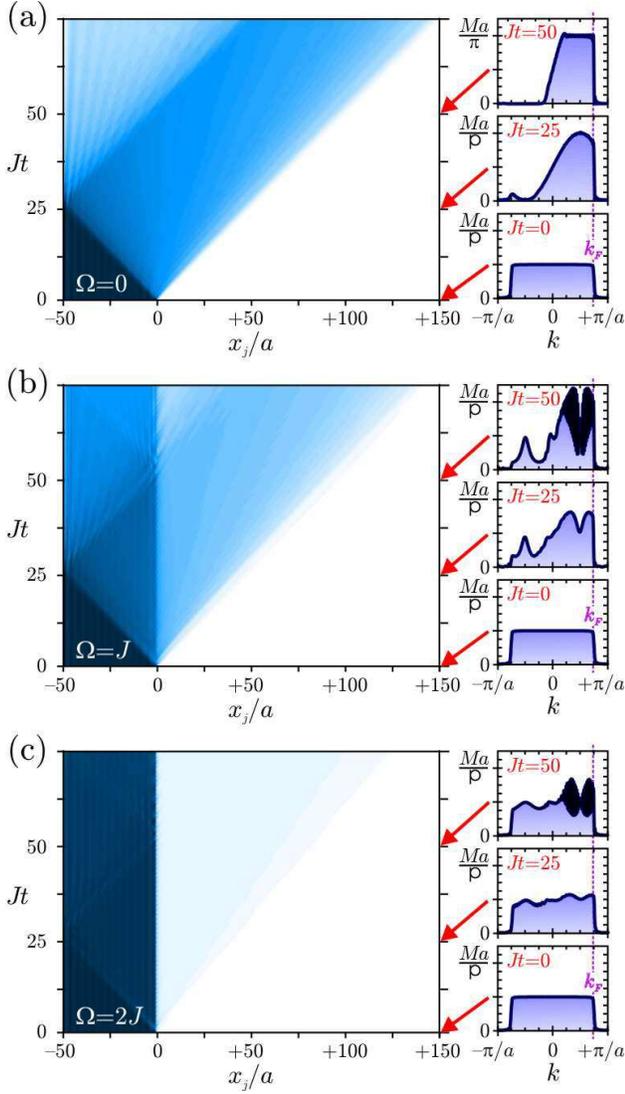}
  \end{center}
  \caption{Evolution of a Fermi-gas with filling factor $\nu=1/2$
    ($N=25$ particles on $M=50$ sites) after opening the switch at
    $t=0$. The effective Rabi-frequency is (a) $\Omega=0$, (b) $\Omega=J$ (c) $\Omega=2J$, and in
    all three cases we have $W=\Delta=0$. The left side shows the
    atomic density $n(x,t)$ (darker regions correspond to higher
    density), and the right side the momentum distribution $n(k,t)$ at
    time $t=0,M/2J,M/J$ (from bottom to top). The corresponding
    density distribution at this three times are indicated by arrow.}
  \label{Fig:FermiDensities}
\end{figure}

We will here detail the time-dependent scattering for a Fermi-gas
of $N$ atoms $A$. For concreteness we assume the fermions to be
initially trapped in a box of $M$ sites to the left the impurity
atom $Q$. The corresponding wavefunction of the system is given by
\begin{eqnarray}
|\Psi(t=0)\rangle = \prod_{n=1}^N \left[\sum_j \alpha_j(k_n) a_j^\dag\right]
 |Q\rangle,\\
\alpha_j(k_n) = \sqrt{\frac{2}{M}}{\Big \{} \begin{array}{cc}\sin(k_n
  x_j)&\mbox{for\enspace} -M{\leq}j\leq1,\\0&\mbox{else},\end{array}\nonumber
\end{eqnarray}
where the quasi-momenta $k_n=n\pi/(M+1)a$. This corresponds to a Fermi-sea
filled up to $E_F=4J\sin^2(k_F a)$, where $k_F=\nu \pi/a$ is the
Fermi-momentum and the initial filling-factor $\nu=N/(M+1)$.\\
At time $t=0$ we open the impurity (cf.~Fig.~\ref{Fig:setup}(b)), and from
Eq.~\eqref{Eq:Fermi_Spin_Model} we obtain the
evolution of the system as
\begin{subequations}
\begin{gather}
|\Psi(t)\rangle = \prod_{n=1}^N\left[\sum_j\alpha_j(k_n,t)a_j^\dag+
\beta(k_n,t)f^\dag\right]|Q\rangle,\label{Eq:Psi_Fermionic_t}\\
\alpha_j(k_n,t) = \sum_{j'}U_{j,j'}(t)\alpha_{j'}(k_n),\label{Eq:alpha_fermionic}\\
\beta(k_n,t) =\sum_{j'}U_{M,j'}(t)\alpha_{j'}(k_n)\label{Eq:beta_fermionic},
\end{gather}
\end{subequations}
where $U_{\alpha,j}(t)$ are the single-particle propagators, cf. Eq.~\eqref{Eq:Propagator_Full}.  According to
Eq.~\eqref{Eq:Psi_Fermionic_t} the scattering from the impurity occurs
independently for each particle in the initial Fermi sea, with
scattering amplitudes $t_k$ and $r_k$ given in Eq.~\eqref{Eq:tk_rk} for $0<k\leq k_F$. The
atomic and molecular densities are thus given by the sum of the
probabilities for the single fermions in the Fermi-gas,
\begin{subequations}
\begin{eqnarray}
n(x_j,t) &=& \langle a_j^\dag a_j \rangle_t =
\sum_{n=1}^N |\alpha_j(k_n,t)|^2,\label{Eq:n_xj_t_fermionic}\\
n_M(t) &=& 1-n_Q(t) = \sum_{n=1}^N |\beta(k_n,t)|^2.
\end{eqnarray}
\end{subequations}
Moreover, after opening the switch the atomic quasi-momentum distribution in the Fermi-gas for
the semi-infinite system is given by
\begin{eqnarray}
n(k,t) &=& \frac{a}{2\pi} \sum_{j,j'}e^{-ik(x_{j'}-x_j)}\langle
a_{j'}^\dag a_j^{\phantom{\dag}}\rangle_t\nonumber\\
&=& \frac{a}{2\pi}\sum_{n=1}^{N} \Big{|}\sum_j
e^{-ikx_j}\alpha_j(k_n,t)\Big{|}^2.
\end{eqnarray}

In Fig.~\ref{Fig:FermiDensities}(a,b,c) we show the evolution for a Fermi-sea with $\nu\approx 3/4$, i.e. $N=38$
particles initially on $M=50$ sites. For each simulation we have
$W=\Delta=0$, but the driving varies as $\Omega/J=0,1,2$ in Fig.~\ref{Fig:FermiDensities}(a,b,c)
respectively. On the left side we plot the atomic density
$n(x_j,t)$ (darker regions correspond to higher density). To the right
we plot the respective momentum profiles $n(k,t)$ for the Fermi-gas at
times $t=0,M/2J,M/J$ (from bottom to top). This times are indicated
by arrows each figure.

In Fig.~\ref{Fig:FermiDensities}(a) we see the evolution of the
noninteracting system, $\Omega=0$.  The atomic cloud expands freely to
the right after opening the switch at $t=0$. The corresponding momentum
distribution is initially given by $n(k,t=0)\approx \theta(k_F-|k|)Ma/2\pi$
(see profile at the bottom). With progressing time $0<t<M/2J$ the gas
develops a forward peak at $k=\pi/2a$ (see profile in the middle) until
becoming asymmetric as $n(k,t=M/J)\approx
(M/2\pi)\theta(k_F-k)2\theta(k)$ for $t\leq M/J$ (see profile at the
top).

In Fig.~\ref{Fig:FermiDensities}(b) we show the behavior for for
weak laser driving, $\Omega=J$. We notice that there is already substantial blocking by the
impurity. The corresponding momentum profiles show that the blocking is
mainly due to the complete reflection of fermions with quasi-momentum
near $k=\pi/2a$ to $k'=-\pi/2a$.

In Fig.~\ref{Fig:FermiDensities}(c) we plot the densities for resonant
driving with $\Omega=2J$. The transport through the impurity is
efficiently blocked by the impurity atom, as the initial densities
$n(x_j,t=0)$ and $n(k,t=0)$ are almost completely preserved.

In the following we will consider the number of particles on the right
side of the impurity,
\begin{eqnarray}
N_R(t) &=& \sum_{j>0} n(x_j,t),
\end{eqnarray}
and the corresponding particle current through the
impurity, $I_R$.
\begin{eqnarray}
I_R(t) &=& \frac{dN_R}{dt}(t).
\end{eqnarray}
They characterize the behavior of the switch.

\begin{figure}[tb]
  \begin{center}
    \includegraphics[width=0.95\columnwidth]{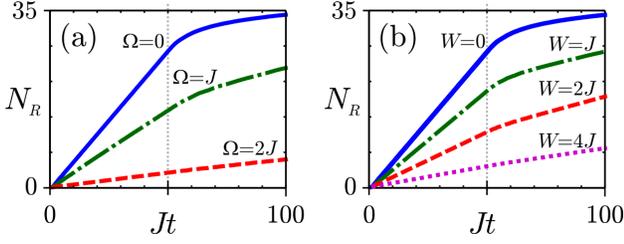}
  \end{center}
  \caption{The number of atoms $A$ transmitted through the impurity as
    a function of time $t$,
    $N_R(t)$, for a Fermi-gas with $\nu=1$. (a) We gave
    $\Delta=W=0$ and the coupling is $\Omega=0$ (solid line), $\Omega=J$
    (dashed-dotted line), $\Omega=2J$ (dashed line). In (b) there is
    no coupling, $\Omega=0$, but the background interaction are
    $W=0,J,2J,4J$ (solid, dash-dotted, dashed, dotted line). The
    system establishes a constant particle current up to $t\approx 50/J$
    $\Delta=W=0$ and the coupling is $\Omega=0$ (solid line),
    $\Omega=M/J=50/J$, and saturates until all $N=50$ atoms $A$
    passed the impurity.}
  \label{Fig:FermiNumbers}
\end{figure}

In Fig.~\ref{Fig:FermiNumbers}(a) we show the number of particles
$N_R(t)$, for the same parameters as in
Fig.~\ref{Fig:FermiDensities}, i.e. for each the initial filling
factor $\nu\approx3/4$ and $W=\Delta=0$. The solid line
shows $N_R$ for the no coupling to the impurity, $\Omega=0$, and corresponds
to the densities shown in Fig.~\ref{Fig:FermiDensities}(a). The
dashed-dotted line shows the behavior for $\Omega=J$,
cf.~Fig.\ref{Fig:FermiDensities}(b), and the dashed line corresponds to
$\Omega=2J$, i.e.~Fig.\ref{Fig:FermiDensities}(c).
Moreover, in Fig.~\ref{Fig:FermiNumbers}(b) we show the number of particles
$N_R(t)$, for initial filling
factor $\nu\approx3/4$ and $\Omega=\Delta=0$, but for an onsite shift
$W=0,J,2J,4J$ (see solid, dash-dotted, dashed, dotted line), respectively.
In general, after a short transient period, of the order of the inverse tunneling rate $1/J$, the
number of particles on the right side of the impurity increases
linearly with $t$. Thereby the system establishes a roughly constant
flux of particles $I_R$ through the impurity. The flux persists up to
$t\approx M/J$, which is indicated by a vertical dotted line in
Fig.~\ref{Fig:FermiNumbers}(a,b). Then the population on the left side of
the impurity is substantially depleted and therefore $N_R(t)$
saturates until all particles tunneled through the impurity, yielding
$N_R(t)=N$ and $I_R(t)=0$ for $t\rightarrow\infty$.

We are interested in the linear regime. From Eq.~\eqref{Eq:n_xj_t_fermionic} we obtain the constant
average current as (cf. the Landauer-formula)
\begin{eqnarray}
I_R = \frac{dN_R}{dt} = \frac{1}{2\pi} \int_0^{k_F} dk v_k T_k,\label{Eq:Fermi_Current}
\end{eqnarray}
where $v_k=-2Ja\sin(ka)$ is the group-velocity of the quasi-particles
with quasimomentum $k$, $k_F=\nu\pi/a=N\pi/(M+1)a$ the
Fermi-quasimomentum and $T_k$ the corresponding transmission
coefficients, cf.~Eq.~\eqref{Eq:tk_rk}. Thus the average current is
obtained by integrating the Fano-profiles $T_k=1-R_k$, see e.g.~Fig~\ref{Fig:Reflection}, up to the
Fermi-momentum.

\begin{figure}[tb]
  \begin{center}
    \includegraphics[width=0.95\columnwidth]{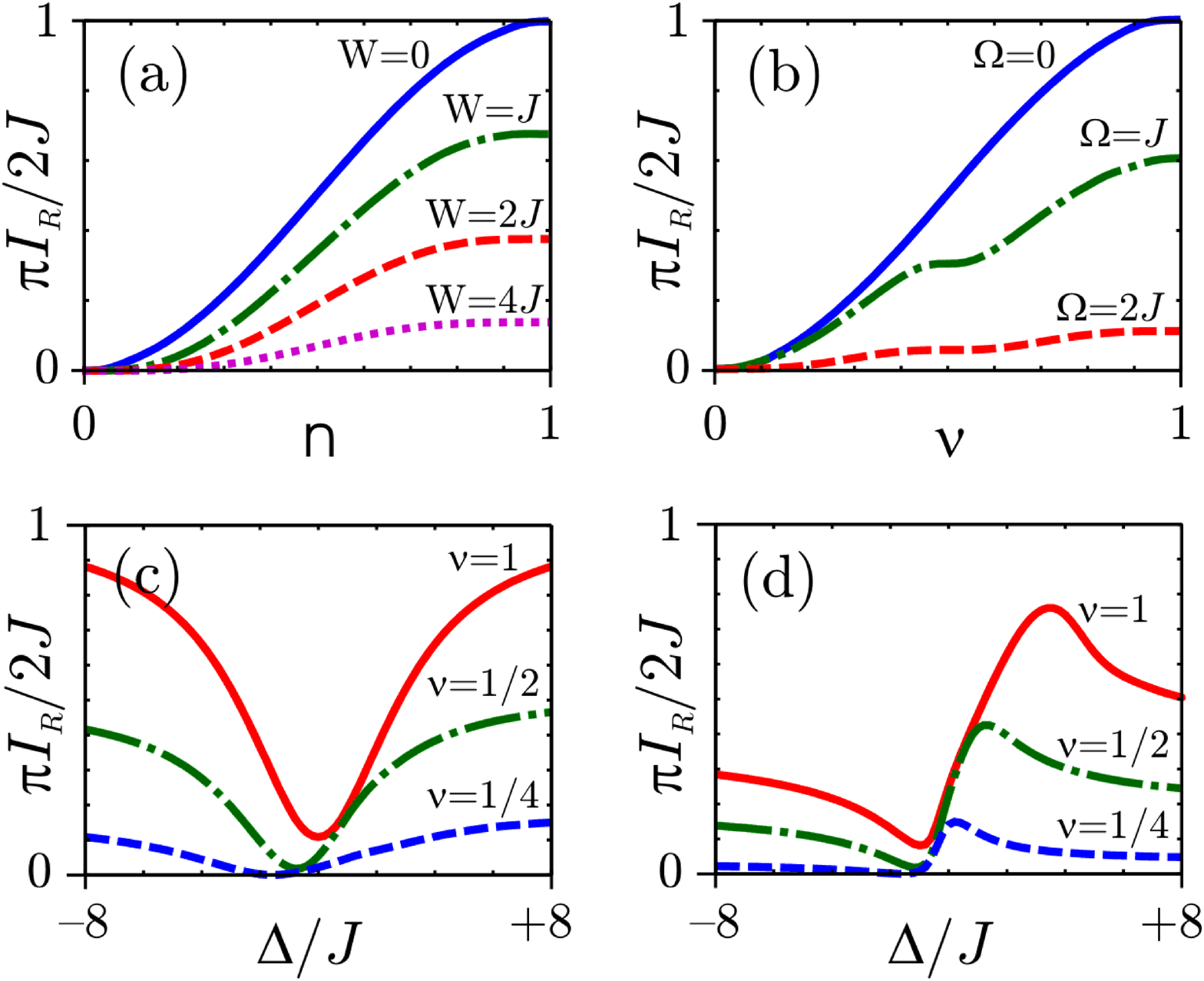}
  \end{center}
  \caption{The quasi-steady state atomic current $I_R$ through the
    impurity. Its dependence on the filling factor $\nu$ is shown in (a)
    for $\Omega=0$ but $W=0,J,2J,4J$, and in (b) for $W=\Delta=0$ but
    $\Omega=0,J,2J$. The dependence of $I_R$ on the detuning $\Delta$ is shown in (c)
    for $\Omega=2J$, $W=0$ and in (d) for $\Omega=W=2J$.}
  \label{Fig:FermiCurrent}
\end{figure}

For an uncoupled impurity, $W=\Omega=0$, we have $T_k=1$. Thus
the current is given up to a constant by the Fermi-energy,
\begin{eqnarray}
I_R^{(0)}&=&\frac{E_F}{2\pi}=\frac{2J}{\pi}\sin^2\left(\frac{\nu\pi}{2}\right).\label{Eq:Fermi_IR_0}
\end{eqnarray}
In Fig.~\ref{Fig:FermiDensities}(b) we plot the dependence of $I_R^{(0)}$ as a function
of the filling factor $\nu$ as a solid line.

For a finite on-site shift, $|W|>0$, but no laser-driving, $\Omega=0$,
we have $T_k<1$, see Eq.~\eqref{Eq:Rk_for_W} and
Fig.~\ref{Fig:Reflection}(a). Thus the current through the switch decreases as
\begin{eqnarray}
I_R^{(W)}&=&\frac{E_F}{2\pi}-\frac{W^2}{2\pi E_0}{\rm
  arccoth}\left(\frac{W^2+2JE_F}{E_0E_F}\right)\label{Eq:Fermi_IR_W}
\end{eqnarray}
where $E_0=\sqrt{W^2+4J^2}$ is the modulus of the bound-state
energy. The exponential decay of $I_R^{(W)}$ with increasing coupling
strength $W$, i.e. the arccoth-term in Eq.~\eqref{Eq:Fermi_IR_W}, is
characteristic for a system with one bound-state. The dependence of the
current $I_R^{(W)}$ on the filling-factor $\nu$ is shown in
Fig.~\ref{Fig:FermiCurrent}(a). The solid line shows the
non-interacting value, $W=0$, while the dash-dotted, dashed, dotted line correspond to
$W=J,2J,4J$, respectively.

However, for resonant driving at $\Omega>0$ and couplings $W=\Delta=0$
we obtain a symmetric Fano-profile for $R_k$ with respect to $k=\pi/2a$, see
Fig.~\ref{Fig:Reflection}(b). Therefore, by integrating the latter profiles
we obtain the current as
\begin{eqnarray}
I_R^{(\Omega)} &=&
\frac{E_F}{2\pi}
-\frac{E_+E_-^2}{2\pi\left(E_+^2-E_-^2\right)}{\rm
    arccoth}\left(\frac{E_-^2+2JE_F}{E_+E_F}\right)+\nonumber\\
&&-\frac{E_+^2E_-}{2\pi\left(E_+^2-E_-^2\right)}{\rm
    arccot}\left(\frac{E_+^2+2JE_F}{E_-E_F}\right),\label{Eq:Fermi_IR_Omega}
\end{eqnarray}
with $E_\pm=\sqrt{\sqrt{\Omega^4+4J^4}\pm 2J^2}$. The arccoth-term in
Eq.~\eqref{Eq:Fermi_IR_Omega} gives the mean effect of the reflection
as that typical of a system with one bound-state, while the oscillating
arccot-term is induced by the presence of two interfering poles in the
scattering matrix. The current
Eq.~\eqref{Eq:Fermi_IR_Omega} is plotted in
Fig.~\ref{Fig:FermiCurrent}(b) as a function of the initial filling
factor $\nu$ for the uncoupled impurity, $\Omega=0$ (solid line),
 for $\Omega=J$ (dashed-dotted line), and for $\Omega=2J$
(dashed line). For finite driving the current $I_R^{(\Omega)}$ shows a plateau at
$\nu=1/2$, as all the Bloch-waves near $k=\pi/2a$ are completely reflected from
the impurity (see~Fig.~\ref{Fig:Reflection}(b)). From
Eq.~\eqref{Eq:Fermi_IR_Omega} we obtain that already for $\Omega\sim
4J$ the current of particles through the impurity is completely
suppressed for arbitrary filling $\nu$, i.e. up to Fermi-energy $E_F=4J$.

In the following we discuss the dependence of the current for
$\Omega>0$ on the detuning $\Delta$. In Fig.~\ref{Fig:FermiCurrent}(c) we show the current $I_R$ for
$\Omega=2J$ but still $W=0$ as a function of the detuning $\Delta$ for
several initial densities $\nu$. The solid line corresponds to
commensurate initial filling, $\nu=1$, the dash-dotted line to
half-filling, $\nu=1/2$, and the dotted line to a dilute Fermi-gas
with $\nu=1/4$. The current shows a symmetric profile with a minimum
at $\Delta\approx-2J\cos^3(\nu\pi/2)$ and approaches its threshold
value $I_R^{(0)}$ for $|\Delta|\gg\Omega$.
Notice that the resonance for the many-body Fermi-gas with increasing
density from the bottom of the Bloch-band, $\Delta\approx-2J$, toward
the middle of the band, $\Delta=0$.

For finite $W=2J$ (see~Fig.~\ref{Fig:FermiCurrent}(d)) the dependence
shows an asymmetric profile and reaches its threshold
value $I_R^{(W)}$ (cf.~Eq.~\eqref{Eq:Fermi_IR_W}) for large detuning, $|\Delta|\gg\Omega^2/|W|$.
We notice that although the single fermions in the Fermi-sea scatter
independently, we obtain a finite current for
$\Omega<J$, even on resonance. This is caused by the fact that the
various fermionic modes $a_k$ see the resonance
 a different (energy-dependent) detuning $\Delta-E_k$, which leads to a shift of the minimum (and maximum) of
the transmitted current proportional to the density of the Fermi-gas,
see Fig.~\ref{Fig:FermiCurrent}(c,d). However, in the limit of strong
driving $\Omega\gg J$ we recover the features of perfect blocking
for $\Delta\approx0$ and of perfect transmission for $\Delta\approx \Omega^2/W$.

\subsection{Ideal Bose-gas}

We now consider the case, where the probe atoms $A$ are spin-less
non-interacting bosons. The Hamiltonian for the system is given by
\begin{eqnarray}
H &=& -J \sum_j \left(a_j^\dag a_{j+1} + {\rm h.c.}\right) + \Delta |M\rangle\langle M| + \nonumber\\
&& + W_Q |Q\rangle\langle Q| a_0^\dag a_0+ W_M |M\rangle\langle M|a_0^\dag a_0+\nonumber\\
&& + \Omega \left( |M\rangle\langle Q| a_0 + a_0^\dag |Q\rangle\langle M| \right),\label{Eq:Hamiltonian_Boson_Basis}
\end{eqnarray}
where the operators $a_j^\dag$ ($a_j$) create (annihilate) an atom $A$
on site $j$, and obey the canonical commutation relations
$[a_i,a_j^\dag]=\delta_{ij}$ and $[a_i,a_j]=[a_i^\dag,a_j^\dag]=0$.
Moreover, $|Q\rangle$ ($|M\rangle$) denote the states with an atom $Q$
(a molecule in state $M$) on the impurity, and $W_Q$ ($W_M$) is the
onsite-shift for an atom $A$ and an atom $Q$ (a molecule $M$) on the
impurity. As in the previous section we will henceforth restrict ourselves to the
case of equal on-site shifts $W_M=W_Q\equiv W$. In this case we may
rewrite the Hamiltonian as
\begin{eqnarray}
H &=& -J\sum_j \left(a_j^\dag a_{j+1} + a_{j+1}^\dag a_j\right) + W
a_0^\dag a_0 +\nonumber\\
&&+\Delta \sigma^+ \sigma^-+\Omega\left(\sigma^+ a_0 + a_0^\dag \sigma^-\right),\label{Eq:Bose_Spin_Model}
\end{eqnarray}
where the pauli operators $\sigma^+\equiv |M\rangle\langle Q|$ and
$\sigma^-\equiv |Q\rangle\langle M|$ obey canonical
anti-commutation relations and commute with $a_j$ and $a_j^\dag$.
The Hamiltonian~\eqref{Eq:Bose_Spin_Model} corresponds a multi-mode  Jaynes-Cummings
model.

In the following we will consider the scattering of a gaussian
wavepacket of $N$ bosons $A$, all initially occupying the same
single particle state, $\alpha(x_j,t=0)$, approaching the impurity
atom $Q$ with mean quasi-momentum $\pi/a>k_0>0$ and width ${\delta
k}_0\ll\pi/a$. The corresponding wavefunction for the system is
given by
\begin{eqnarray}
|\Psi(t=0)\rangle &=& |Q\rangle\frac{1}{\sqrt{N!}}\left[\sum_j \alpha(x_j,t=0)
 a_j^\dag\right]^N |{\rm vac}\rangle,\\
\alpha_0(x_j,0)&=&{\cal N}e^{-{\delta k}_0^2(x_j- x(0))+i k_0 x_j},
\end{eqnarray}
where for $\delta
k_0\ll\pi/a$ the normalization is given by ${\cal N}^2=(2\delta
k_0^2a^2/\pi)^{1/2}=n_0/N$ in terms of the peak density of the
gaussian wavepacket, $n_0\equiv n(x(0),t=0)$, and $x(0)\ll 0$ denotes the mean position of the particles $A$ at $t=0$.

For $\Omega=0$ the equations of motion for $a_j$ decouple from
$\sigma^-$. Therefore, we obtain the scattering of the
bosons $A$ by the impurity, as
\begin{eqnarray}
|\Psi(t)\rangle &=& |Q\rangle\frac{1}{\sqrt{N!}}\left[\sum_j \alpha(x_j,t)
 a_j^\dag\right]^N |{\rm vac},
\end{eqnarray}
where the single-particle wavefunction for finite $W$ was already
obtained in Sec.~\ref{Sec:Single_Particle_Scattering}. For this
case all the results obtained in
Sec.~\ref{Sec:Single_Particle_Scattering} hold, and we obtain e.g.
the density as $N$ times the single particle result,
 $n(x_j,t)=N|\alpha(x_j,t)|^2|$.

\subsubsection{Linearization of the impurity}

For $\Omega\gg J \gg |\Delta|,|W|$, we obtained in Sec.~III that
the population of the molecular state was strongly suppressed,
i.e. as $(J/\Omega)^2$, and thus we approximate
$\sigma^z\rightarrow-1$. Thus we linearize the spin, i.e. set
$\sigma^+\rightarrow b^\dag$ and $\sigma^-\rightarrow b$, where
$b$ and $b^\dag$ obey canonical commutation relations. The
scattering of the bosons $A$ by the impurity is given by
\begin{eqnarray}
|\Psi(t)\rangle &\approx& \frac{1}{\sqrt{N!}}\left[\sum_j \alpha_j(t)
 a_j^\dag + \beta(t)b^\dag\right]^N |{\rm vac}\rangle_A,\label{Eq:Psi_Bose_Approx}
\end{eqnarray}
where the single-particle wavefunction $\alpha_j(t)$ and the amplitude $\beta(t)$
of the molecular state $b^\dag|{\rm vac}\rangle\equiv\sigma^+|Q\rangle=|M\rangle$,
 were obtained in Sec.~\ref{Sec:Single_Particle_Scattering} (see Eq.~\eqref{Eq:Propagator_Full}).

\begin{figure}[tb]
  \begin{center}
    \includegraphics[width=0.95\columnwidth]{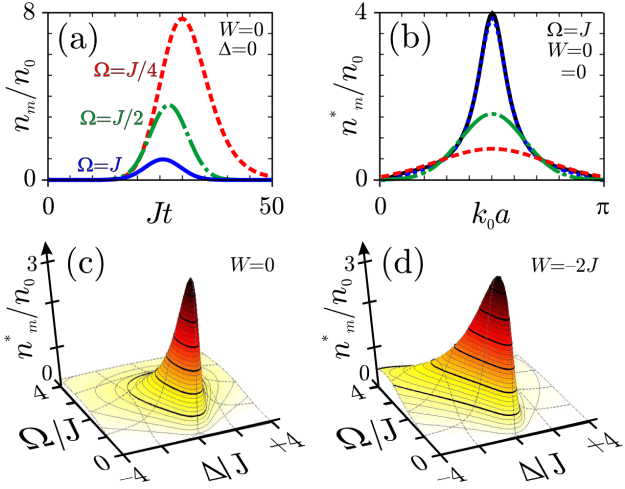}
  \end{center}
  \caption{Scattering
    of a Gaussian wavepacket of bosonic atoms $B$ with mean quasi-momentum
    $k_0$ and width ${\delta k}_0$ as obtained by linearizing the
    impurity (see text). (a) Molecular population as a function of
    time for a narrow wavepacket with $k=\pi/a$ and $\delta
    k_0=\pi/100a$ for $\Omega=J,J/2,J/4$ (solid, dash-dotted, dashed
    line). (a) The maximal attained molecular population ,
    $n_m^\star$, during the
    interaction process with a weakly coupled impurity ($\Omega=J$, $\Delta=W=0$), as a function of the
    quasi-momentum $k_0$ of the incoming wavepacket. The three lines
    correspond a wavepacket with width $\delta k_0=\pi/200a,\pi/50a,\pi/20a$ (solid,
    dash-dotted and dashed line). The dependence of $n_m^\star$ on the
    parameters $\Omega$ and $\Delta$ for $k=\pi/2a$ and $\delta
    k_0=\pi/50a$ in (c) without the presence of background collisions,
    $W=0$, and in (d) for an onsite-shift $W=2J$. }
 \label{Fig:Bose_Approx}
\end{figure}

Self-consistency of the replacement $\sigma_z\rightarrow -1$ requires that the obtained
molecular population $n_m(t)\ll 1$. From the linearization we obtain the molecular population as (see App.~\ref{App:Gaussian})
\begin{eqnarray}
n_m(t) &=& \frac{n_0}{4{\delta k}_0^2\pi}\left|\int dk
\frac{\Omega t_k e^{-i E_k t}}{E_k-\Delta}e^{-\tilde k^2/4{\delta k}_0^2-i \tilde k x(0)}\right|^2,\label{Eq:BoseApprox_nm_of_t}
\end{eqnarray}
where $\tilde k\equiv k-k_0$, and the Fourier-integral is obtained analytically e.g. by using a saddle-point
method, see App.~\ref{App:Gaussian}. We find that
the maximal attained molecular population, $n_m^\star$, is
proportional to the initial density $n_0$ of the gas. In the case of a
broad resonance, $\Omega> \delta k_0$, we find
\begin{eqnarray}
n_m^* &\approx& \frac{n_0\Omega^2v_0^2/a^2}{\left(E_0-\Delta\right)^2v_0^2/a^2+\left[\Omega^2+W\left(E_0-\Delta\right)^2\right]^2},\label{Eq:BoseApprox_nm_max1}
\end{eqnarray}
with $E_0=-2J\cos(k_0a)$ and $v_0=2J\sin(k_0 a)$. Moreover, for a
extremely narrow resonance, $\Omega<\delta k_0$ and $|\Delta|<J$,
one obtains
\begin{eqnarray}
n_m^* &\approx& \frac{n_0}{2\delta
  k_0^2a}\frac{\Omega^2}{W^2+v_\star^2/a^2}e^{-\left(k_0-k_\star\right)^2/2\delta
k_0^2},\label{Eq:BoseApprox_nm_max2}
\end{eqnarray}
where $v_\star=2J\sin(k_\star a)$ and $k_\star\approx\arccos(-\Delta/2J)$ denotes the position of the maximum of
$T_k/(E_k-\Delta)^2$.

In Fig.~\ref{Fig:Bose_Approx}(a) we show the molecular population, $n_m$, as obtained by the replacement
$\sigma_z\rightarrow -1$. In Fig.~\ref{Fig:Bose_Approx}(a) we plot the
$n_m(t)/n_0$ for an incoming gaussian wavepacket with $k_0=\pi/2a$ and
$\delta k_0=0.01\pi/a$ for driving $\Omega=J/4$ (dashed line), $\Omega=J/2$
(dashed-dotted line) and $\Omega=J$ (solid line). In all three cases
we have $\Delta=W=0$. We see that with increasing Rabi-frequency
$\Omega$ the attained molecular-population quickly drop as $\sim
J^2/\Omega^2$, and that the molecular-population closely resembles the
density distribution $n(x_j,t=0)$ of the atomic-cloud $A$. In
Fig.~\ref{Fig:Bose_Approx}(b) we plot the maximal attained population,
$n_m^*$, as a function of the incoming momentum of the gas, $k_0$, for
$\Omega=J$ and $W=\Delta=0$. The four lines correspond to different
width $\delta k_0$ of the wavepacket, i.e. $\delta k_0=0.001\pi/a$ (solid line),
$\delta k_0=0.01\pi/a$ (dashed line), $\delta k_0=0.1\pi/a$
(dash-dotted line) and $\delta k_0=0.2\pi/a$ (dashed line). For a
given width $\delta k_0$ the
molecular population attains its maximum for $k_0=\pi/2a$, i.e. where
we have complete reflection of the wave-packet (see
Fig.~\ref{Fig:Reflection}(b)). At the point of
complete-reflection, $k_0=\pi/2a$ the
population attains its overall maximum for a narrow
momentum-distribution, i.e. for $\delta k_0\rightarrow 0$ we have
$n_m\approx 4n_0$ for $\Omega=J$ and $\Delta=W=0$.
The dependence of $n_m^\star$ on the detuning $\Delta$ and the
Rabi-frequency $\Omega$ is shown in Fig.~\ref{Fig:Bose_Approx}(c,d)
for $W=0$ and $W=-2J$, respectively. In both figures the gaussian wavepacket has $k_0=\pi/2a$ and $\delta
k_0=0.02\pi/a$, i.e. initially extends about $\sim\pi/a\delta
k_0=50$ lattice sites. For $W=0$ we have complete reflection of the
wavepacket for $\Delta=0$, and the attained molecular population is
maximal, $n_m\approx3n_0$, for $\Omega\approx2J$ and $\Delta\approx0$
(see Fig.~\ref{Fig:Bose_Approx}(c)).
However, for finite $W=-2J$ we have also complete transmission of the
wavepacket, i.e. for $\Delta=\Omega/W$. From
Fig.~\ref{Fig:Bose_Approx}(d) we notice that for a given $Omega$ the maximal population is
shifted from $\Delta\approx0$ towards the point, where one has complete
transmission of the wavepacket, i.e. $\Delta\approx
\Omega/W$. However, the overall maximum of $n_m^\star$ in both cases,
$W=0$ and $W=-2J$, is attained for $\Omega\approx2J$ and for stronger driving quickly
drops as $n_m^\star\approx n_0 (2J/\Omega)^2$.
As the replacement $\sigma^-\rightarrow b$ is self-consistent for a dilute
gas with densities $n_0\ll(n_m^\star/n_0)^{-1}$, we see from
\ref{Fig:Bose_Approx} that the approximation holds, even on resonance
$\Delta=0$, for strong driving $\Omega>4J$ up to densities as high as
$n_0\sim 5$ and for small densities $n_0<1$ only fails for $\Delta\approx 0$ and
$\Omega\approx 2J$.

\subsubsection{Time-dependent Variational Ansatz}

In the following we use a time-dependent variational Ansatz to
describe the behavior of the many-body wavefunction in near resonance
$\Delta\approx 0$ for $\Omega\sim J$, i.e. in the regime where the
approximation $\sigma_z\rightarrow-1$ fails already for small
densities, $n_0>0.05$.  As a generalization of
Eq.~\eqref{Eq:Single_Ansatz} for $N\geq1$ bosonic atoms $A$ we choose
as an number-conserving Ansatz for the
state of the system
\begin{eqnarray}
|\Psi(t)\rangle &=&
  c_Q(t)|Q\rangle\frac{\left[a_Q(t)^\dag\right]^N}{\sqrt{N!}}|{\rm vac}\rangle + \nonumber\\
&& c_M(t)|M\rangle\frac{\left[a_M(t)^\dag\right]^{N-1}}{\sqrt{({N-1})!}}|{\rm vac}\rangle,\label{Eq:Bose_Variational_Ansatz}
\end{eqnarray}
where $a_\sigma(t)=\sum_j \alpha_{j,\sigma}(t) a_j^\dag$ represent
two non-orthogonal time-dependent modes for the field of the
bosonic atoms $A$ given that the impurity is in state
$\sigma=Q,M$, and the amplitudes for the impurity, $c_\sigma(t)$,
and for the bosonic wavepackets, $\alpha_{j,\sigma}(t)$, are
normalized as
$\sum_\sigma|c_\sigma(t)|^2=\sum_j|\alpha_{j,\sigma}(t)|^2$. The
equation of motion for variational parameters, $c_\sigma(t)$ and
$\alpha_{j,\sigma}(t)$, are obtained by minimizing the
corresponding action (cf.~App.~\ref{App:Variational})
\begin{eqnarray}
S(t) &=& \frac{\langle\dot\Psi_t|\Psi_t\rangle-\langle\Psi_t|\dot\Psi_t\rangle}{2i}-\langle\Psi_t|H|\Psi_t\rangle\label{Eq:Bose_Variational_Action}
\end{eqnarray}
with respect to $c_\sigma(t)^*$ and $\alpha_{j,\sigma}(t)^*$, as a
set of coupled non-linear differential equations, which we
integrate numerically. Thus we obtain the dynamics of the system.

\begin{figure}[tb]
  \begin{center}
    \includegraphics[width=0.95\columnwidth]{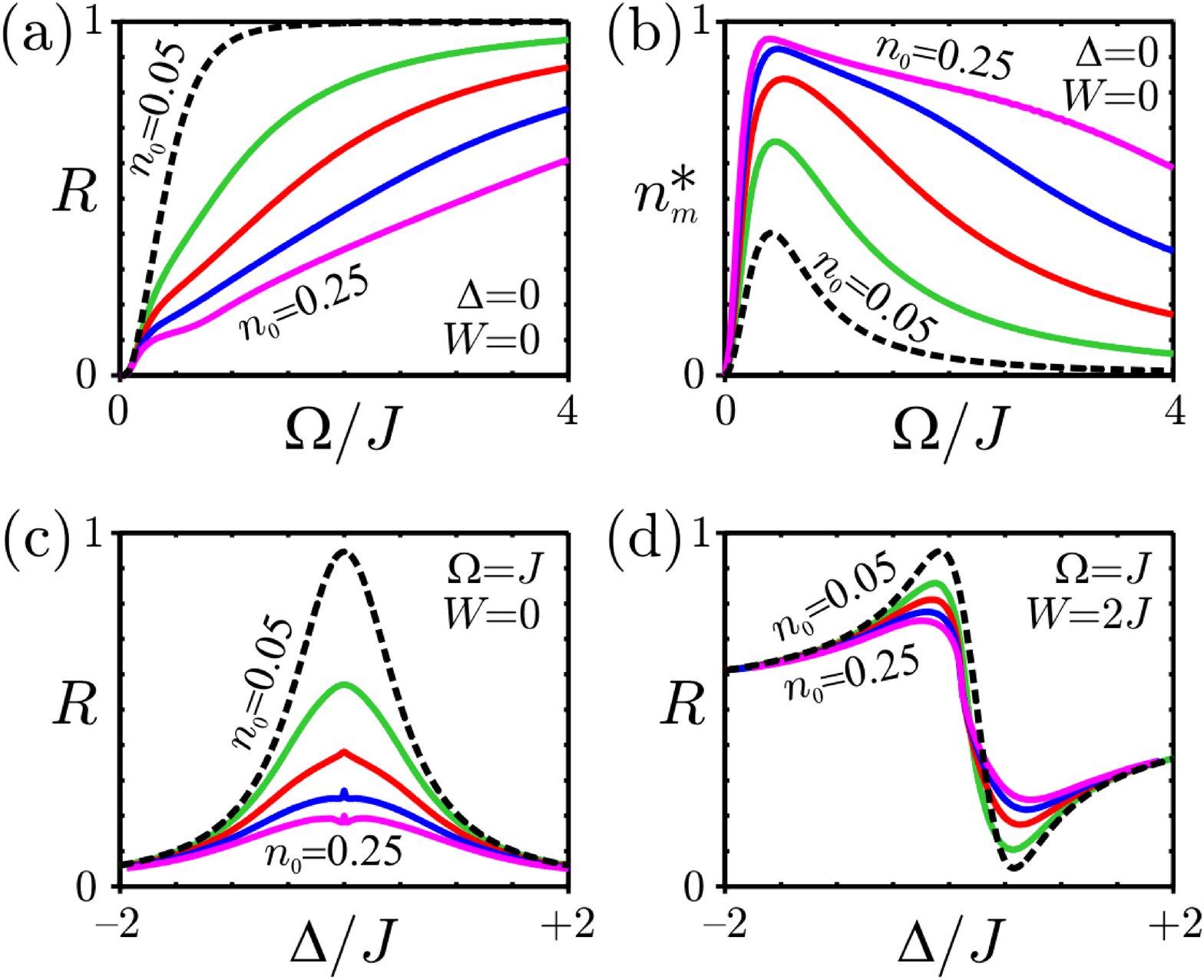}
  \end{center}
  \caption{Evolution of a Gaussian wavepacket with mean quasi-momentum
    $k=\pi/2a$ and width $\delta k_0=0.02\pi/a$ obtained from the variational
    Ansatz Eq.~\eqref{Eq:Bose_Variational_Ansatz}. In (a) the reflection
    coefficient $R$ is plotted as a function of the Rabi-frequency
    $\Omega$ for $\Delta=W=0$, and (b) shows the corresponding attained maximal
    molecular population $n_m^\star$ during the interaction with the
    impurity atom. The reflection coefficient as a function of the
    detuning $\Delta$ for $\Omega=J$ and (c) $W=0$ and (d) $W=2J$. The
    various solid lines correspond to initial densities
    $n_0=0.1,0.15,0.20,0.25$. The dotted-line is for $n=0.05$ and corresponds to
    a linearization of the impurity, see Eq.~\eqref{Eq:Psi_Bose_Approx}.}
 \label{Fig:Bose_Variational}
\end{figure}

In Fig.~\ref{Fig:Bose_Variational} we show the obtained
reflection-coefficient
\begin{eqnarray}
  R=\lim_{t\rightarrow\infty}\frac{1}{N}\sum_{j<0}n(x_j,t)
\end{eqnarray}
and the attained peak molecular population $n_m^\star$ for a
Gaussian-wavepacket with narrow momentum $k_0=\pi/2a$, i.e.
$\delta k_0=0.02\pi/a\ll\pi/a$ for initial density
$n_0=0.05,0.1,0.15,0.20,0.25$ (values indicated in plots).  In
Fig.~\ref{Fig:Bose_Variational}(a)
(Fig.~\ref{Fig:Bose_Variational}(b)) we plot $R$ ($n_m^\star$) as
a function of the Rabi frequency $\Omega$ for $\Delta=W=0$, i.e.
when complete reflection of the wavepacket was predicted by the
bosononic approximation of the spin. The reflection shows a
non-linear behavior in the density for $\Omega>.1$, i.e. decreases
as $\sim 1/n_0$ with increasing density $n_0$. While for
$n_0=0.05$ (see dotted line) we have complete reflection of the
wavepacket for $\Omega>2J$, for higher densities we still have a
finite transmission at $\Omega=2J$.  From
Fig.~\ref{Fig:Bose_Variational}(a) we see that with increasing
density the transmission coefficient rapidly deviates from the
low(zero)-density result and approaches a linear behavior in
$\Omega$ already for $n_0=1/4$. In
Fig.~\ref{Fig:Bose_Variational}(a) we plot the dependence of
$n_m^\star$ on $\Omega$, which shows that the maximal population
is attained for $\Omega\approx J/2$ and  decreases as $\sim
(J/\Omega)^2$ for $\Omega\gg J n_0$. Moreover, we see that with
increasing density, $n_0>0.10$, the peak in the molecular
population, $n_m^\star$, is no longer linear in the density as was
predicted by linearization, cf.~Eq.~\eqref{Eq:BoseApprox_nm_max1}
and cf.~Eq.~\eqref{Eq:BoseApprox_nm_max1}, but saturates toward
the unitary limit $n_m^\star\approx1$.  The dependence of the
reflection coefficient $R$ on the detuning $\Delta$ is plotted in
Fig.~\ref{Fig:Bose_Variational}(c) for $\Omega=J$ and $W=0$ and in
Fig.~\ref{Fig:Bose_Variational}(c) for $\Omega=J$ and $W=2J$. In
the limit of a very dilute Bose-gas (see dashed lines for
$n_0=0.05$) we obtain the single-particle result $T=|t(k_0)|^2$
given by Eq.~\eqref{Eq:tk_rk}, showing a symmetric Fano-profile
for $W=0$ and an asymmetric Fano-profile for $W=2J$, (see also
Fig.~\ref{Fig:Reflection}(b,d)). We notice that at such
weak-driving as $\Omega=J$ the peak (and the asymmetry) in the
Fano-profiles are suppressed with increasing density. However, for
strong driving $\Omega\gg 4J n_0$ we recover the features of
complete reflection (complete transmission) through the impurity
site as was already predicted by the linearization of the impurity.

\subsection{Hard-core Bosons}

We now consider the limit of a strongly interacting Bose-gas. Its Hamiltonian is given by
\begin{eqnarray}
H &=& -J \sum_{j}\left( a_{j+1}^\dag a_j + {\rm h.c.}\right) + \frac{U}{2} \sum_j a_j^\dag a_j^\dag a_j a_j + \nonumber\\
&& + \Delta |M\rangle\langle M| + \Omega \left( |M\rangle\langle Q| a_0
  + a_0^\dag|Q\rangle\langle M| \right)+\nonumber\\
&&+      W_Q a_0^\dag |Q\rangle\langle Q| a_0 + W_M a_0^\dag a_0 |M\rangle\langle
      M|\label{Eq:Hamiltonian_Boson_Basis_U}
\end{eqnarray}
where the onsite-shift for two-bosons $A$ on the same site, $U$,
by far exceeds the tunneling rate $J$, i.e. $U{\gg}J$. Since
double occupation of a site by two atoms $A$ is strongly
suppressed, we may eliminate those excitation from $H$, e.g. by
imposing $a_j^2\equiv0$. In the following we will focus on the
limiting case $U/J\rightarrow\infty$, i.e. that of a Tonks gas. In
this limit we may fermionize the Hamiltonian
\eqref{Eq:Hamiltonian_Boson_Basis_U} via a Jordan-Wigner
transformation (JWT) \cite{Sachdev}, which maps the commuting
fields for the hard-core bosons, $a_j$, and the pseudo-spin of the
impurity, $\sigma^-\equiv|M\rangle\langle Q|$, onto anticommuting
fields, $c_j$ and $f$, respectively. The JWT is given by
\begin{subequations}
\begin{eqnarray}
a_j = c_j\prod_{l<j}\left(1-2c_l^\dag c_l\right),
a_j^\dag = c_j^\dag\prod_{l<j}\left(1-2c_l^\dag
  c_l\right),\\
\sigma^- = f\prod_{l}\left(1-2c_l^\dag
  c_l\right),
\sigma^+ = f^\dag\prod_{l}\left(1-2c_l^\dag c_l\right).
\end{eqnarray}\label{Eq:Jordan_Wigner_Transformation}
\end{subequations}
The fields $c_j$ and $f$ describe fermionic excitations for the new
joint vacuum state of the system, $|{\rm vac}\rangle_{CF}\equiv|Q\rangle|{\rm
  vac}\rangle$. We rewrite the Hamiltonian
\eqref{Eq:Hamiltonian_Boson_Basis_U} in terms of the fermionic
excitations, $c_j$ and $f$, and obtain
\begin{eqnarray}
H &=& -J \sum_{j<M} \left( c_{j+1}^\dag c_j + {\rm h.c.}\right) + \nonumber\\
&& + \Delta f^\dag f + \Omega \left(-1\right)^{\hat N_R}\left( f^\dag c_0
 + {\rm h.c.} \right)+\nonumber\\
&&+      W_Q f f^\dag c_0^\dag c_0 + W_M f^\dag f c_0^\dag c_0,\label{Eq:Hamiltonian_HardCoreBoson_Fermionic}
\end{eqnarray}
where $\hat N_R=\sum_{j>0} c_j^\dag c_j=\sum_{j>0} a_j^\dag a_j$ is the
number of particles to the right of the impurity site. The Hamiltonian
 for the fermionic excitations, $c_j$ and $f$, is the same as the
one obtained for the Fermi-gas, cf.~Eq.~\eqref{Eq:Fermi_Spin_Model},
except for the appearance of the phase-factor $(-1)^{\hat N_R}$ for the
coupling $\Omega$ to the impurity.

We proceed by detailing the time-dependent scattering of a tonks
gas with $N$ atoms $A$ off the impurity atom $Q$. We assume that at
time $t=0$ the atoms $A$ are trapped within a box of $M$ sites to the
left on impurity site. This corresponds to a fermi-sea of the
fermionic modes $c_j$, and the state of the system at $t=0$ is given
by (see also Sec.\ref{Sec:Ideal_Fermi_gas})
\begin{gather}
|\Psi(0)\rangle = \prod_{n=1}^N \left[\sqrt{\frac{2}{M}}\sum_{j<0}
 \sin\left(k_n x_j\right)c_j^\dag\right]|{\rm
 vac}\rangle_{CF}\nonumber\\
= |Q\rangle\sum_{{\bf j}<0} (-1)^{S({\bf j})}\prod_{n=1}^N \left[\sqrt{\frac{2}{M}}
 \sin\left(k_n x_{j_n}\right)a_{j_n}^\dag\right]|{\rm vac}\rangle_{A},
\end{gather}
where $k_n=n\pi/(M+1)a$ are the quasi-momenta of the fermionic
excitations, ${\bf j}=(j_1,j_2,\ldots,j_N)$ (with $j_p\neq j_q$)
denotes the position of the $N$
bosons and $(-1)^{S({\bf j})}$ the permutational sign of ${\bf
j}$, i.e. $S({\bf j})=\sum_{j_p>j_q}1$. Due to the cumbersomeness of
the many-body wavefunction in terms of the bosonic operators $a_j$, it is preferable to deal within the fermionic
picture and extract the quantities of interest from the correlations
for the fermions. The density $n(x_j,t)$ of the hardcore bosons $A$ corresponds
to the density for the fermions $c$, while the correlations,
$\rho(x_i,x_j,t)$, and the momentum distribution of the
Tonks gas, $n(k,t)$, differ from those of a Fermi-gas, as
\begin{subequations}
\begin{eqnarray}
n(x_j,t) &=& \langle a_j^\dag a_j \rangle = \langle c_j^\dag c_j \rangle,\\
\rho(x_i,x_j,t) &=& \langle a_i^\dag a_j \rangle = \langle c_i^\dag
\prod_{l=i}^j\left[1-2c_{l}^{\dag}c_{l}\right] c_j \rangle,\\
n(k,t) &=& \frac{a}{2\pi}\sum_{m,n}e^{-ik\left(x_{m}-x_n\right)}\langle a_{m}^\dag
a_n\rangle\\\label{Eq:Fermi_correlations}
&=& \frac{a}{2\pi}\sum_{m,n}e^{-ik\left(x_{m}-x_n\right)}\langle c_{m}^\dag
\prod_{l=m}^n \left[1-2c_l^\dag c_l\right]c_n\rangle.\nonumber
\end{eqnarray}
\end{subequations}
Diagonalizing the single-particle density matrix $\rho(x_i,x_j,t)$ one obtain the
condensate fraction $N_0(t)$ as the largest eigenvalue and the
wavefunction of the quasi-condensate $\psi_0(x_j,t)$ as the corresponding eigenmode. In the following we
denote density of the quasi-condensate as $n_0(x_j,t)$ and its momentum distribution as $n_0(k,t)$.

\begin{figure}[thb]
  \begin{center}
    \includegraphics[width=0.8\columnwidth]{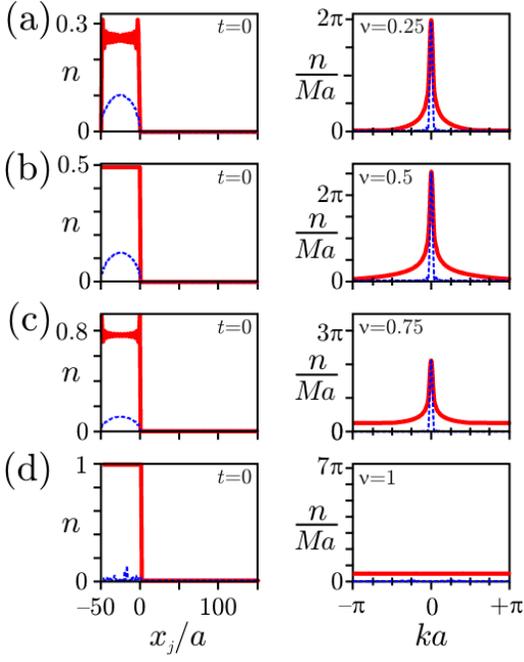}
  \end{center}
  \caption{Initial density distribution of a gas of hard core bosons. The gas is trapped on
    $M=50$ to the left of the impurity site, $x_j=0$, and the
    fillings factor is (a) $\nu=1/4$, (b) $\nu=1/2$, (c) $\nu=3/4$ and
    (d) $\nu=1$, respectively. Shown are the atomic density $n$ (solid
    line) and the mode of the quasi-condensate $n_0$ (dotted line), in position
    (left column) and momentum space (right column).}
  \label{Fig:HardCoreBoseDensities_initial}
\end{figure}

In Fig.~\ref{Fig:HardCoreBoseDensities_initial}
we plot the initial density for a Tonks-gas trapped on $M=50$ sites
for various filling factors $\nu=1/4,1/2,3/4,1$, i.e. we have $N=13,25,38,50$ particles for
Fig.~\ref{Fig:HardCoreBoseDensities_initial}(a,b,c,d),
respectively. The solid lines in the plot on the left shows the
density in position space, $n(x_j)$, and the dotted lines show the
contribution of the largest eigenmode of the
single-particle-density matrix $\rho(x_i,x_j)$, $n_0(x_j)$. To the right we plot
the corresponding quasi-momentum distributions of the gas, $n(k)$ (solid line), and
 for the largest eigenmode of $\rho(x_i,x_j)$, $n_0(k)$ (dotted line).
While the density merely resembles that of a homogeneous Fermi-gas
with local filling factor $\nu$, the momentum distributions
strongly differs from the typical Fermi-sea (cf.
Fig.~\ref{Fig:FermiDensities}) as it shows a sharp peak at $k=0$,
as one would expect from a condensate. However, the condensed
fraction $N_0\equiv\sum_j n_0(x_j)$ (see dashed lines within the
same figures) is not macroscopic, as it amounts only to
$\approx\sqrt{N}$ particles in the gas and thus the behavior of
the Tonks-gas significantly differs from that of a true BEC. In
fact we notice that the momentum distribution shows considerable
wings, which account for the depletion in the quasi-condensate.
Moreover, with increasing density $\nu$ the number of particle in
the quasi-condensate depletes considerably until vanishing
completely for $\nu=1$ (see
Fig.~\ref{Fig:HardCoreBoseDensities_initial}(d)), where the system
attains a Mott-insulator with no phase-correlations.

\begin{figure}[thb]
  \begin{center}
    \includegraphics[width=0.8\columnwidth]{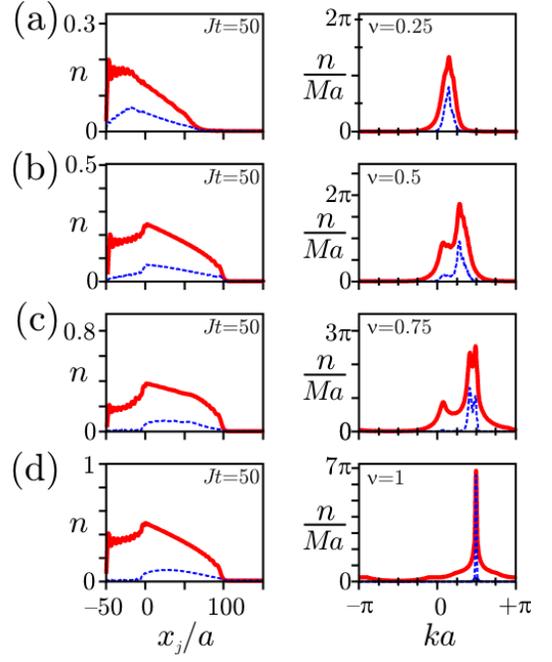}
  \end{center}
  \caption{Free evolution of a gas of hard core bosons. The gas was
    initially trapped on $M=50$ to the left of the impurity site, $x_j=0$, and the
    fillings factor is (a) $\nu=1/4$, (b) $\nu=1/2$, (c) $\nu=3/4$ and
    (d) $\nu=1$, respectively. Shown are the atomic density $n$ (solid
    line) and the density of the quasi-condensate $n_0$ (dotted line), in position
    (left column) and momentum space (right column) at a time $t=M/J$
    after opening the mirror.}
  \label{Fig:HardCoreBoseDensities_final}
\end{figure}

We now detail the free ($\Omega=W_Q=0$) evolution of the Tonks
gas after having opened the switch at $t=0$. From
Eq.~\eqref{Eq:Hamiltonian_HardCoreBoson_Fermionic} we obtain the state
of the system at time $t$ to be given by (cf. Sec.\ref{Sec:Ideal_Fermi_gas})
\begin{subequations}
\begin{eqnarray}
|\Psi_t\rangle &=&
 |Q\rangle\prod_{n=1}^N\left[\alpha_j(k_n,t)c_j^\dag\right]|{\rm
 vac}\rangle,\\
\alpha_j(k_n,t)&=& \sqrt{\frac{2}{M}}\sum_{j'<0} U_{j,j'}(t)\sin(k_n x_j),
\end{eqnarray}\label{Eq:HCB_Psi_t_fermionic}
\end{subequations}
where $U_{j,j'}(t)$ denotes the free single-particle propagator,
cf.~Eq.~\eqref{Eq:Propagator_Full} with $W=\Omega=0$. The density
$n(x_j,t)$ of the hardcore-bosons $A$ corresponds to that of the
fermions, given by Eq.~\eqref{Eq:n_xj_t_fermionic}.  In
Fig.~\ref{Fig:HardCoreBoseDensities_final} we show to the left the
densities $n(x,t)$ of the Tonks-gas (solid lines) and of the
condensate mode $n_0(x_j,t)$ (dashed line) at time $t=M/J=50/J$
after opening the switch. To the right we plot the corresponding
momentum distributions, $n(k,t)$ and $n_0(k,t)$. As in
Fig.~\ref{Fig:HardCoreBoseDensities_initial} the subplots
Fig.~\ref{Fig:HardCoreBoseDensities_final}(a,b,c,d) correspond to
an initial filling factor $\nu=1/4,1/2,3/4,1$, respectively. The
density of the Tonks-gas, $n(x_j,t)$, corresponds to the one
obtained for a Fermi-gas (see
e.g.~Fig.~\ref{Fig:FermiDensities}(a) for $\nu=1/2$), and shows
the spreading of the gas through the impurity. The corresponding
momentum distribution of the Tonks-gas, $n(k,t)$, is shifted away
from $k=0$ towards $k>0$ and spread in momentum space, due to the
tunneling of the particles through the impurity site. Moreover,
after a brief transient period starts developing an new additional
peak at $k\approx k_F/2=\nu\pi/2$, which gains in magnitude until
reaching its maximum value at $t=M/J$.  This corresponds to the
dynamical formation of a quasi-condensate $n_0(k,t)$ \cite{Rigol},
which now propagates as a wave-packet through the impurity with
$k\approx\nu\pi/2$ (see dotted lines in
Fig.~\ref{Fig:HardCoreBoseDensities_final}).  The number of
particle is the quasi-condensate, $N_0(t)=\sum_j n_0(x_j,t)$,
gives the magnitude of the peak and grows with progressing time
until it saturates at $t\approx M/J$. At this time the Tonks gas
on the right side of the impurity is significantly depleted and
one has $N_0(t)\approx\sqrt{N}$.  The dynamical formation of a
quasi-condensate propagating with $k>0$ is clearly seen for
filling factors $\nu>1/2$, where it momentum peak at $k=k_F/2$
exceeds the magnitude of the initial $k=0$ coherent fraction. For
the Mott-insulating state (commensurate filling $\nu=1$) the
situation is particularly clear: the mott-insulator melts through
the impurity (see left side of
Fig.~\ref{Fig:HardCoreBoseDensities_final}(d)) and forms a
quasi-condensate propagating as a coherent wavepacket with
velocity $v_k=2Ja$ through the lattice. Thereby the initially flat
develops a sharp peak at $k=\pi/a$ (see right side of
Fig.~\ref{Fig:HardCoreBoseDensities_final}(d)). The number of
particles in the quasi-condensate increases until reaching
$\sim\sqrt{N}$ for $t=M/J$.  It saturates as the initial
Mott-insulating state is significantly depleted. Moreover, from
the evolution of the density distributions in position space (see
plots on the right of
Fig.~\ref{Fig:HardCoreBoseDensities_final}(a-d)) we see that,
although the quasi-condensate only amounts to a fraction of atoms
in the system, its profile $n_0(x_j,t)$ describes the
\emph{transmitted} part of the Tonks-gas through the impurity
accurately, i.e. up to a multiplicative factor.

\begin{figure}
\begin{center}
\includegraphics[width=0.95\columnwidth]{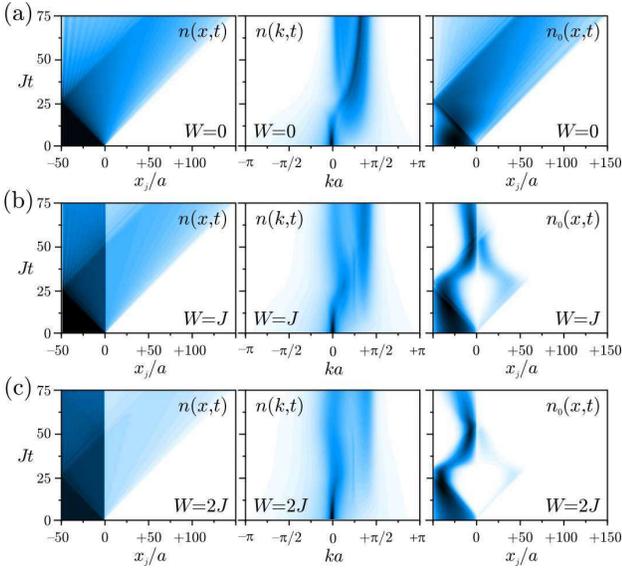}
\caption{Evolution of a gas of hard-core bosons with filling factor
  $\nu=1/2$ for $\Omega=0$ and (a) $W=0$, (b) $W=J$, (c) $W=2J$. Shown
  are the density $n(x_j,t)$ (left column), the momentum
  distribution $n(k,t)$ and the density of the quasi-condensate
  $n_0(x_j,t)$.}\label{Fig:HCB_dens_W}
\end{center}
\end{figure}

In the presence of a finite on-site shift, $W_Q\neq 0$, but with
driving $\Omega=0$, the Hamiltonian is still bilinear in the
fermionic modes $c_i^\dag$ and $c_j$. The evolution of the system
is given by Eq.~\eqref{Eq:HCB_Psi_t_fermionic}, but now with
$U_{j',j}(t)$ being the single-particle propagator for $W\neq0$.
In Fig.~\ref{Fig:HCB_dens_W}(a,b,c) we show the evolution of a
Tonks gas with $\nu=1/2$ for $W=0$, $W=J$ and $W=2J$,
respectively. On the left side we plot the density distribution,
$n(x_j,t)$, in the middle the momentum distribution, $n(k,t)$, and
to the right the quasi-condensate $n_0(x_j,t)$. Darker regions
correspond to regions of higher density. For no interaction,
$W=0$, the particles freely tunnel through the impurity.  The
momentum distribution is shifted towards $k>0$ as well as being
broadened. At $t\approx M/2J=25/M$ an additional peak in the
momentum distribution is formed at $k\approx\pi/3a$, which
corresponds to the mode of the quasi-condensate, $n_0(x_j,t)$,
tunneling through the impurity (see Fig.~\ref{Fig:HCB_dens_W}(a)).
For $W=J$ the tunneling of particles through the impurity site is
partially blocked and the dynamical formation of a monochromatic
mode with $k>0$ is suppressed, as is see from the. Thereby the
condensate mode $n_0(x_j,t)$ remains mainly localize to the left
side of the impurity. For $W=2J$ we see that only a small fraction
of atoms passes through the impurity and the momentum distribution
remains centered at $k=0$, although becoming slighty broader. We
also see that the condensed fraction of atoms in the condensate is
efficiently hindered from passing the impurity and remains
localized to the left of the impurity site.

In the presence of a finite coupling $\Omega>0$, the Hamiltonian
is no more bilinear in $c_j^\dag$ and $c_j$, due the appearance of
the nonlinear factor $(-1)^{\hat N_R}=\prod_{j>0}(1-2c_j^\dag
c_j)$, which makes a description of the time-dependent scattering
in terms of the fermionic modes $c_j$ in the general case as
difficult as integrating out the full many-body Schr\"odinger
equation \eqref{Eq:Hamiltonian_Boson_Basis_U} for the hard-core
bosons $A$. However, the contribution of the nonlinear factor
$(-1)^{\hat N_R}$ to the dynamics is negligible for strong
driving, $\Omega \gg J,|\Delta|,W$, and also for low densities,
$\nu\ll 1$. In this case the number of atoms on the right
$N_R(t)\ll 1$, and those we set $\hat N_R\rightarrow 0$ in
Eq.~\eqref{Eq:Hamiltonian_HardCoreBoson_Fermionic}. The state of
the system is given by
\begin{subequations}
\begin{eqnarray}
|\Psi_t\rangle &=&
 \prod_{n=1}^N\left[\alpha_j(k_n,t)c_j^\dag+\beta(k_n,t)f^\dag\right]|{\rm
 vac}\rangle,\\
\alpha_i(k_n,t)&=& \sqrt{\frac{2}{M}}\sum_{j<0} U_{i,j}(t)\sin(k_n
 x_j),\\
\beta(k_n,t)&=& \sqrt{\frac{2}{M}}\sum_{j<0} U_{M,j}(t)\sin(k_n x_j),
\end{eqnarray}\label{Eq:HCB_Psi_t_fermionic_Omega_approx}
\end{subequations}
where $U_{\alpha,j}(t)$ denotes the full single-particle propagator,
cf.~Eq.~\eqref{Eq:Propagator_Full}. In this regime the density
distribution of the Tonks-gas, $n(x_j,t)$, corresponds to that of a
Fermi-gas (see e.g. the left side in Fig.~\ref{Fig:FermiDensities}).

For the general case of many bosons $A$, arbitrary coupling strengths,
and even finite $U$, we refer to the exact numerical simulation given
in Ref.~\cite{Daley}. These simulations allow to test the behavior of the gas for
essentially arbitrary repulsion $U$ and density $\nu$, i.e. for the full crossover regime from
a weakly interacting dilute Bose-gas up to a dense Tonks gas.

\section{Conclusion}

  We have studied a scheme utilizing quantum interference to control the
  transport of atoms in a 1D optical lattice by a single impurity
  atom. The two internal state represent a qubit (spin-1/2), which in
  one spin state is perfectly transparent to the lattice gas, and in
  the other spin state acts as a single atom mirror, confining the
  lattice gas. This allows to ``amplify'' the state of the qubit, and
  provides a single-shot quantum non-demolition measurement of the
  state of the qubit.  We have derived exact analytical expression for the
  scattering of a single atom by the impurity, and gave approximate
  expressions for the dynamics a gas of many interacting bosonic of
  fermionic atoms.

  A numerical study of this dynamics based on
  time-dependent DMRG techniques, which complements the present
  discussion, will be presented in Ref.~\cite{Daley}.

\acknowledgments

The authors acknowledge helpful discussion with A.~J.~Daley and D.~Jaksch.
Work in Innsbruck is supported by the Austrian Science Foundation, EU
Networks, and the Institute for Quantum Information.

\appendix

\section{Scattering of Gaussian Wavepackets}\label{App:Gaussian}

We consider the dynamics of a bosonic $N$-particle state of the form
\begin{eqnarray}
  |\Psi_t\rangle &=& \frac{1}{\sqrt{N!}}\left[\sum_j \alpha_t(x_j)
    a_j^\dag + \beta_t b^\dag\right]^N|{\rm vac}\rangle,\nonumber
\end{eqnarray}
where at $t=0$ the gaussian wave-packet is given by $\beta_0=0$ and
\begin{eqnarray}
  \alpha_0(x_j)&=&{\cal N}_0 e^{-{\delta k}_0^2
    \left(x_j-x_0\right)+i k_0 x_j},\nonumber
\end{eqnarray}
where $x_0\ll0$ ($k_0>0$) is the mean position (momentum) and ${\delta
  k}_0<\pi/a$ the width in momentum space at $t=0$. For ${\delta
  k}_0\ll\pi/a$ we may take the continuum limit $\sum_j\rightarrow\int
dx/a$ and obtain ${\cal N}_0 \approx \left(2{\delta
    k}_0^2a^2/\pi\right)^{1/4}$, and ${\delta x}_0\approx 1/2{\delta
  k}_0\gg a/\pi$.
The momentum representation of the is given by
\begin{eqnarray}
\tilde\alpha_0(k)&=&\left(a/2\pi\right)^{1/2}\sum_j e^{-i*k*x_j}
\alpha_0(x_j)\nonumber\\
&\approx&\left(2\pi{\delta k}_0^2\right)^{-1/4}e^{-\left(k-k_0\right)^2/4{\delta k}_0^2-i\left(k-k_0\right)x_0}.\nonumber
\end{eqnarray}

\subsection{Molecular density}

For the evolution we are interested at the population of the molecular
state, which follows as
\begin{eqnarray}
n_m(t)&=&\langle b^\dag b \rangle_t = N \left|\sum_j U_{M,j}(t)
  \alpha_0(x_j)\right|^2\nonumber\\
&=&N{\cal N}_0^2\left|\sum_j U_{M,j}(t) e^{-{\delta k}_0^2
 \left(x_j-x_0\right)+i k_0 x_j}\right|^2,\nonumber
\end{eqnarray}
where $U_{M,j}(t)=\langle M | e^{-i H t} | j \rangle$ is the single
particle propagator. We introduce the peak of the initial atomic
density $n_0\equiv n(x_0,t=0) = N{\cal N}_0^2\approx N\left(2/\pi\right)^{1/2}{\delta k}_0a$.

For the scattering off the particles, i.e. for $|x_0|\gg r_B$, we may
neglect the finite range of the bound-state and have
\begin{eqnarray}
n_m(t)&\approx&N\left|\sum_j \frac{a}{2\pi}\int dk e^{-iE_k t}\frac{\Omega
    t_k}{E_k-\Delta} e^{i k |x_j|} \alpha_0(x_j) \right|^2\nonumber\\
&=&N\left(\frac{a}{2\pi}\right)\left|\int dk e^{-iE_k t}\frac{\Omega
    t_k}{E_k-\Delta} \tilde\alpha_0(k) \right|^2.\nonumber
\end{eqnarray}

Thus we have
\begin{eqnarray}
n_m(t)&=&n_0 \left|f_t\right|^2,\nonumber\\
f_t&=&\frac{1}{2{\delta k}_0\sqrt{\pi}}\int dk
\frac{\Omega t_k}{E_k-\Delta}e^{-i E_k t-\left(\frac{k-k_0}{2{\delta k}_0}\right)^2-i\left(k-k_0\right)x_0}.\nonumber
\end{eqnarray}

For the Fourier transform we use a saddle-point method, however, we
have to distinguish which one is narrower, either the width of the wave-packet,
$\delta k_0$, or the width of dressing profile.

\subsection{Broad Fano-profile}

For a broad resonance, i.e. $\Omega t_k/(E_k-\Delta)$ slowly
varying on the Bloch band we expand the integral around $k\approx
k_0$, i.e. with
\begin{eqnarray}
E_k \approx E_0 + v_0 \left(k-k_0\right) +
\frac{1}{2m_0}\left(k-k_0\right)^2,\nonumber\\
\frac{\Omega t_k}{E_k-\Delta}\approx \frac{\Omega t_0}{E_0-\Delta},\nonumber
\end{eqnarray}
where energy $E_0 \equiv E_k\mid_{k=k_0}$, velocity $v_0 \equiv
\partial E_k/\partial k\mid_{k=k_0}$ and effective mass $m_0 \equiv 1/(\partial^2
E_k/\partial^2 k)\mid_{k=k_0}$. Notice that $m_k=-1/a^2E_k$ for the explicit shape of the Bloch-band and
choice of the origin in the band-middle.

Thereby we obtain
\begin{eqnarray}
n_m(t)&=&n_0 \frac{\delta k_t}{\delta k_0}e^{-2\delta k_t x_t^2}
{\cal D}_0,\nonumber
\end{eqnarray}
where the linear propagation, spreading and the dressing are
\begin{eqnarray}
x_t &=& x_0 + v_0 t,\nonumber\\
\delta k_t &=& \frac{\delta k_0}{\sqrt{1+\left(2\delta k_0^2 a^2 E_0
      t\right)^2}},\nonumber\\
{\cal D}_0 &=&
\frac{\Omega^2|t_k|^2}{\left(E_k-\Delta\right)^2}\mid_{k=k_0}
\nonumber\\
&=&
\frac{\Omega^2}{\left(E_0-\Delta\right)^2+\left[\Omega^2+W\left(E_0-\Delta\right)\right]^2a^2/v_0^2}.\nonumber
\end{eqnarray}

We notice that at $t_\star=-x_0/v_0=|x_0/v_0|$ we attain a maximal molecular
density of
\begin{eqnarray}
n_m(t_\star)&=&n_0 \frac{{\cal D}_0}{\sqrt{1+\left(2\delta k_0^2 a^2 E_0
      t\right)^2}}.\nonumber
\end{eqnarray}
For $E_0\approx 0$ we might neglect the broadening/spreading.
We recognize that ${\cal D_0}$ is maximal for such detuning $\Delta$
and initial momentum $k_0$ where
\begin{eqnarray}
E_\star &\approx& \Delta - W \frac{\Omega^2}{W^2+v_\star^2/a^2},\nonumber
\end{eqnarray}
is on the bloch-band. This corresponds to the position of the
Fano-profile. As $v_\star^2/a^2=4J^2-E_\star^2$ the Equation for
$E_\star$ is implicitly cubic gives the same recursive equation.
Thus the maximal density for a broad resonance $\Omega\gg |J-\Delta|$ is
suppressed as
\begin{eqnarray}
n_m(t_\star) \approx n_0 \frac{v_0^2/a^2}{\Omega^2},
\end{eqnarray}
while far off resonance $|\Delta|\gg\Omega$ we have
\begin{eqnarray}
n_m(t_\star) \approx n_0 \frac{(\Omega/\Delta)^2}{1+(aW/v_0)^2}.
\end{eqnarray}

\subsection{Narrow Fano-profile}

In the second case, that of a narrow resonance, i.e. a sharp Fano
profile, we have that the dressing factor is narrower than the
Gaussian wavepacket, hence we approximate via a Saddle-point method.
We expand the dressing function,
\begin{eqnarray}
{\cal D}_k = \frac{\Omega
    t_k}{E_k-\Delta} = \left[\frac{ia\Omega}{v_k}+\frac{E_k-\Delta}{\Omega}\left(1+\frac{iaW}{v_k}\right)\right]^{-1},\nonumber
\end{eqnarray}
around the momentum $k_\star$ where $|{\cal D}_k|$ is maximal,
\begin{eqnarray}
{\cal D}_k \approx C_0 e^{+i C_1 (k-k\star)a -\frac{1}{2}C_2 (k-k_\star)^2a^2},\label{Eq:AppExpansion}
\end{eqnarray}
with
\begin{eqnarray}
C_0 &=& \frac{-i\Omega}{\zeta},\nonumber\\
C_1 &=& \frac{\gamma_\star+iW}{\zeta}-\frac{i m_\star}{\gamma_\star}\left(1+i\frac{E_\star-\Delta}{\zeta}\right),\nonumber\\
C_2 &=& \left(\frac{\gamma_\star+iW}{\zeta}\right)^2
        -\frac{im_\star}{\gamma_\star}\left(\frac{\gamma+iW}{\zeta}+\frac{2\Omega^2}{\zeta^2}\right)\nonumber\\
        &&+\frac{m_\star^2}{\gamma_\star^2}\left[1+\frac{\left(E_\star-\Delta\right)^2}{\zeta^2}\right],\nonumber\\
\zeta &\equiv&
\frac{\Omega^2+\left(E_\star-\Delta\right)\left(W-i\gamma_\star\right)}{\gamma_\star},\nonumber
\end{eqnarray}
where $E_\star=E(k_\star)$, $\gamma_\star=\partial E(k)/\partial k
a\mid_{k=k_\star} = v(k_\star)/a$, $m_\star=\partial^2 E(k)/\partial (k
a)^2 = -E(k_\star)$ are the lowest expansion coefficient of the
dispersion relation $E(k)=-2J\cos(ka)$ at $k=k_\star$.

The position of its maximum $k_\star$ is obtained from the expansion~\eqref{Eq:AppExpansion}by requiring $\Im[C_1]=0$ with
\begin{eqnarray}
\Im[C_1]=
\gamma_\star\frac{\Omega^2W+\left(E_\star-\Delta\right)\left(\gamma_\star^2+W^2\right)}
     {\Omega^4+2\left(E_\star-\Delta\right)\Omega^2W+\left(E_\star-\Delta\right)\left(\gamma_\star^2+W^2\right)}\nonumber\\
- \frac{m_\star}{\gamma_\star}\frac{\left[\Omega^2+\left(E_\star-\Delta\right)W\right]^2}
     {\Omega^4+2\left(E_\star-\Delta\right)\Omega^2W+\left(E_\star-\Delta\right)\left(\gamma_\star^2+W^2\right)}.\nonumber
\end{eqnarray}
In the limit of interest (i.e. near the middle of the Bloch-band), we
have $|m_k/\gamma_k|=|\cot(ka)|\ll 1$ thus we obtain the energy of the
maximum as a series
\begin{eqnarray}
E_\star\approx \Delta - \frac{W\Omega^2}{\gamma_\star^2+W^2}
+
\frac{m_\star\gamma_\star^2\Omega^4}{\left(\gamma_\star^2+W^2\right)^3}
+
{\cal O}^2\left(m_\star\right).
\end{eqnarray}
Since $\gamma_\star$, $m_\star=-E_\star$ all depend on $k_\star$ the
series gives implicitly the value of $k_\star$. Moreover, we notice
that the truncation to first order in $m_\star$ yields the exact
result for $W=0$.

Then with
\begin{eqnarray}
\zeta &\approx&
\frac{\Omega^2}{\gamma_\star-iW}\left[1-i\frac{m_\star}{\gamma_\star}\left(\frac{\Omega\gamma_\star}{W^2+\gamma_\star^2}\right)^2\right],\nonumber\\
C_0&\approx&
\frac{-\left(W+i\gamma_\star\right)/\Omega}{1-im_\star\gamma_\star\Omega^2/\left(\gamma_\star^2+W^2\right)^2},\nonumber\\
C_1&\approx& \frac{\gamma_\star^2+W^2}{\Omega^2}-\frac{m_\star
  W}{\gamma_\star^2+W^2},\\
C_2&\approx&
\left(\frac{\gamma_\star^2+W^2}{\Omega^2}\right)^2-i\frac{m_\star}{\gamma_\star}\frac{\gamma_\star^2-W^2-4i\gamma_\star
  W}{\Omega^2},\nonumber
\end{eqnarray}
we can compute the Fourier-integral and obtain
\begin{eqnarray}
|f_t|^2&=&|C_0|^2|A(t)|e^{-\frac{1}{2}\Re\left[A(t)\left(2\delta k_0 x(t) + i
 \frac{k_0-k_\star}{\delta
 k_0}\right)^2+\left(\frac{k_0-k_\star}{\delta
 k_0}\right)^2\right]},\nonumber\\
A(t) &=& \left[1+2\delta k_0^2 a^2\left(C_2+im_\star t\right)\right]^{-1},\\
x(t) &=& x_0+v_\star t-a C_1.\nonumber\\
\end{eqnarray}

For $E_\star\approx 0$, i.e. near the middle of the Bloch-band, we might
neglect the spreading of the wave-packet, i.e. $m_\star\approx 0$.
Then we have
\begin{eqnarray}
|C_0|^2&=&C_1=\frac{W^2+v_\star^2/a^2}{\Omega^2}=\sqrt{C_2},\nonumber\\
|f(t)|^2 &=& \frac{C_1 \exp\left[-\frac{2\delta k_0^2
    x(t)^2+C_1^2\left(k_0-k_\star\right)^2a^2}{1+2\delta k_0^2 a^2
      C_1^2}\right]}{1+2{\delta k_0}^2 a^2 C_1^2}.
\end{eqnarray}
In the limit $\Omega\rightarrow 0^+$ the probability $|f(t)|^2$ vanishes as
\begin{eqnarray}
{\rm max}_t |f(t)|^2 &\approx&
 \frac{\Omega^2}{W^2+v_\star^2/a^2}\frac{\exp\left[-\frac{\left(k_0-k_\star\right)^2}{2\delta
 k_0^2}\right]}{2{\delta k_0}^2a^2}.
\end{eqnarray}

\section{Variational Ansatz}\label{App:Variational}

For the Ansatz~\eqref{Eq:Bose_Variational_Ansatz} we obtain the
action~\eqref{Eq:Bose_Variational_Action} as
\begin{eqnarray}
  \begin{gathered}
    S= \frac{1}{2}\sum_\sigma\Big{[}c_\sigma^*\frac{\partial
      c_\sigma}{\partial t}-\Delta\delta_{\sigma
      M}|c_\sigma|^2+N_\sigma |c_\sigma|^2\sum_j
    \alpha_{j,\sigma}\nonumber\\
    \times\left(i\frac{\partial\alpha_{j,\sigma}}{\partial
        t}+J\alpha_{j+1,\sigma}+J\alpha_{j-1,\sigma}-W\delta_{j0}\alpha_{j,\sigma}\right) \Big{]}\nonumber\\
    -\Omega\sqrt{N}c_M^*c_Q\left(\sum_j
      \alpha_{j,M}^*\alpha_{j,Q}\right)^N\alpha_{j,Q} + {\rm
      c.c.},\nonumber
  \end{gathered}
\end{eqnarray}
with $c_\sigma\equiv c_\sigma(t)$, $\alpha_{j,\sigma}\equiv
\alpha_{j,\sigma}(t)$ and $N_Q=N_M+1=N$.

Minimizing the action $S$ with respect to $c_\sigma^*$ and
$\alpha_{j,\sigma}$, we obtain after some algebra,
\begin{eqnarray}
ic_M^*\frac{\partial c_M}{\partial t} &=& \Delta|c_M|^2 + \lambda -
(N-1)\frac{\lambda+\lambda^*}{2},\nonumber\\
ic_Q^*\frac{\partial c_Q}{\partial t} &=& \lambda^* -
N\frac{\lambda+\lambda^*}{2},\nonumber
\end{eqnarray}
\begin{eqnarray}
i|c_M|^2\frac{\partial \alpha_{j,M}}{\partial t} &=&
|c_M|^2\left(-J\sum_\pm\alpha_{j\pm1,M}+W\delta_{j0}\alpha_{j,M}\right)\nonumber\\
&&-
\frac{\lambda-\lambda^*}{2}\alpha_{j,M}+\frac{\lambda}{s}\alpha_{j,Q},\nonumber\\
i|c_Q|^2\frac{\partial \alpha_{j,Q}}{\partial t} &=&
|c_Q|^2\left(-J\sum_\pm\alpha_{j\pm1,Q}+W\delta_{j0}\alpha_{j,Q}\right)\nonumber\\
&&+
\frac{\lambda-\lambda^*}{2}\alpha_{j,Q}+\frac{N-1}{N}\frac{\lambda^*}{s^*}\alpha_{j,M}
+ \frac{\lambda^*\delta_{j0}}{N\alpha_{0,Q}^*}\nonumber
\end{eqnarray}
where the overlap $s$ and the effective coupling $\lambda$ are given by
\begin{eqnarray}
s &=& \sum_j \alpha_{j,M}^*\alpha_{j,Q},\nonumber\\
\lambda &=& \Omega \sqrt{N} c_M^*c_Q s^{N-1} \alpha_{0,Q}.\nonumber
\end{eqnarray}


\end{document}